\documentstyle[12pt,fullpage,fleqn,psfig]{article}

\newcommand{\be}{\begin{eqnarray}}
\newcommand{\ee}{\end{eqnarray}}
\newcommand{\Tr}{{\rm Tr}}
\renewcommand{\d}{\partial}
\newcommand{\vect}{\left ( \begin{array}{c}}
\newcommand{\evect}{\end{array} \right )}
\newcommand{\mat}{\left ( \begin{array}{cc}}
\newcommand{\emat}{\end{array} \right )}
\renewcommand{\Re}{{\rm Re}}

\begin{document}


\begin{flushright}
SUNY-NTG-00/11
\end{flushright}

\begin{center}
\vskip 2.0cm
{\Large \bf QCD-like Theories at Finite Baryon Density}

\vskip 1.2cm
J.B. Kogut$^1$, M.A. Stephanov$^2$,
D. Toublan$^3$, J.J.M. Verbaarschot$^3$ and A. Zhitnitsky$^4$
\vskip 0.2cm
{\it
$^1$Loomis Laboratory of Physics, University of Illinois,
Urbana-Champaign, IL 61801, USA
\\
$^2$Department of Physics, University of Illinois, Chicago, IL
60607-7059, USA
\\
and
\\
RIKEN-BNL Research Center, Brookhaven National Laboratory, Upton, NY 11973,
USA 
\\
$^3$Department of Physics and Astronomy, SUNY,
Stony Brook, New York 11794, USA
\\
$^4$Department of Physics and Astronomy, University of British Columbia,
Vancouver, BC V6T 1Z1, Canada 
}

\vskip 1.5cm
{\bf Abstract}
\end{center}

We study QCD-like theories with pseudoreal fermions
at finite baryon density.
Such theories include two-color QCD with quarks in the fundamental
representation of the color group as well as any-color QCD with quarks in
the adjoint color representation.
In all such theories the lightest baryons are diquarks.
At zero chemical potential $\mu$ they
are, together with the pseudoscalar mesons, the Goldstone modes
of a spontaneously broken enlarged chiral symmetry group.
Using symmetry principles, we derive the low-energy effective Lagrangian
for these particles. We find that  a second order phase transition
occurs  at a value of $\mu$ equal to half the mass of the Goldstone
modes. For values of $\mu$ beyond this point the scalar diquarks Bose
condense and the diquark condensate is 
nonzero. We calculate the dependence
of the chiral condensate,  the diquark condensate, the baryon charge
density, and the masses of the diquark and pseudoscalar excitations
on $\mu$ at finite bare quark mass and scalar diquark source. 
The relevance of our results to lattice QCD calculations
and to real three-color QCD at finite baryon density is discussed.


\vfill\eject

\pagestyle{plain}

\section{Introduction}

The study of strong interactions at finite baryon density has a long
history. Phenomenological knowledge of nuclear forces allows
one to obtain a good understanding of equilibrium
nuclear matter which is relatively dilute. However, these results
cannot be simply extended to higher densities relevant to neutron stars,
supernova explosions and relativistic heavy ion collisions, where
the microscopic degrees of freedom of QCD, quarks and gluons, 
become important. 

Understanding QCD at finite density has been a tremendous
challenge \cite{Bielefeld,lattice.mu}. Unlike finite temperature QCD, where
significant progress 
has been achieved using lattice Monte Carlo simulations, QCD at finite
baryon density is not amenable to such a numerical approach.
The primary reason is that the determinant of the
Euclidean QCD Dirac operator is not real at finite
baryon chemical potential, $\mu$.

Recent progress has been achieved analytically by studying QCD at
infinite density and by studying models of the Nambu-Jona-Lasinio
type as well as the instanton liquid model \cite{SC1}. It was found that
the QCD vacuum at sufficiently high baryon density could become a color
superconductor \cite{SC1}. In other words, due to the attraction in the
color anti-triplet isosinglet channel the quark Fermi surface becomes
unstable towards the formation of a condensate of diquark pairs.

No first principle lattice calculation methods exist at this moment to
study the phenomenon of color superconductivity. However, the
mechanisms which lead to the formation of diquark condensates (gluon
exchange or instanton induced attraction) can be investigated 
for QCD with two colors. This presents a
tremendous advantage.  QCD with two colors
can be studied numerically on the lattice
\cite{SU2lattice.old,SU2lattice.new} --- 
the determinant of the Dirac operator is real. Another class of QCD-like
theories with diquark condensation that can be studied on
the lattice at nonzero chemical potential is
QCD with $any$ number of colors and
of quarks in the adjoint color representation.
 Clearly, the physics of both types of theories is different from that of the
three-color QCD, but the differences are easy to understand and
classify. We hope that understanding the behavior of such theories at
finite baryon number density will provide us with an additional
insight into the phenomenon of diquark condensation in three-color QCD.
In addition, numerical simulations of both  QCD with two colors and 
QCD with adjoint quarks are now being pursued by
several groups \cite{SU2lattice.new,Thomasmu}. 

The third class of theories which  can be studied in the same way
is QCD with  the phase of the quark determinant quenched. 
In other words, for each quark such theory contains
a conjugate quark, with opposite baryon charge. This leads to the appearance
of colorless diquark states --- baryonic pions. 
The zero-flavor limit $N_f \rightarrow 0$ of such a theory 
is the quenched approximation of QCD, as 
has been demonstrated analytically in \cite{St96} 
using a random matrix theory at nonzero $\mu$.

The unifying property of all such theories is the
pseudo-reality of the quark representations, which manifests itself in
the fact that the determinant of the Dirac operator is real. Another
unifying property (which is ultimately related to
this pseudo-reality) is the fact that diquarks can make color singlets
--- these are the baryons of such theories. The Bose-Einstein condensation of
diquarks, with nonzero baryon charge, 
may be viewed as baryon charge
superconductivity (rather than color superconductivity as in three-color
QCD).

In this paper we construct the low-energy chiral Lagrangian describing
mesons and baryons (diquarks) 
at finite baryon density for two-color QCD and QCD with adjoint quarks.
As was pointed out in \cite{KST} the $\mu$-dependence of this
Lagrangian describing mesons and baryons
can be fixed by global and certain {\em local}
flavor symmetries. Within the domain of validity of this Lagrangian we
study both its vacuum properties and 
the mass spectrum of the Goldstone modes.
Our goal is twofold: (i) to understand and
describe quantitatively the physics associated with diquark
condensation; (ii) to provide lattice theorists with the qualitative
and quantitative predictions aiding their data analysis.

This paper is organized as follows. In section \ref{sec:classes} 
we give an overview 
of the different patterns of chiral symmetry breaking at $\mu=0$
for three-color QCD, two-color QCD and for QCD with adjoint 
quarks. We relate these patterns to the antiunitary 
symmetries of the Dirac operator, which provide us with a convenient
classification in terms of the Dyson index $\beta$.
Section \ref{sec:globsym} reviews the global symmetries of the theories,
with the emphasis on the enlarged $SU(2N_f)$ symmetry. The spontaneous
breaking of this symmetry by the chiral condensate is the
subject of section \ref{sec:breaking}, where we identify the
Goldstone excitations.
The effective Lagrangian at nonzero bare quark mass $m$ is introduced 
in section \ref{sec:baremass}. In section 
\ref{sec:mu} 
we review the introduction of the chemical potential into the
effective Lagrangian by means of
a local gauge principle \cite{KST,mulargeNc}. 
The vacuum alignment dictated by the effective Lagrangian
is analyzed in sections \ref{sec:vacuum} and \ref{sec:globalminimum}.
In section \ref{sec:curv} we expand the static
part of Lagrangian
around the minimum and in section \ref{sec:spec} we determine the masses of
the excitations as functions of $\mu$ and $m$ from the pole of their
respective propagators.
In section \ref{sec:dqsource} we introduce the diquark source $j$ and
find its effect on the vacuum and the mass spectrum.
In section \ref{sec:condensates} we present the dependence of
the vacuum condensates and the baryon number density on $\mu$, $m$ and $j$.
Finally, in section \ref{sec:bose}, 
we rederive the equation of state of the dilute Bose gas of diquarks
with repulsion and show that it exactly matches the equation
of state obtained using our mean field analysis of the previous sections.
Concluding remarks and discussion are presented in section 
\ref{sec:conclusions}.

\section{Overview and classification}
\label{sec:classes}

At zero chemical potential the spontaneous breaking of the chiral
symmetry is believed to be  an essential low-energy property of 
three color QCD with fundamental fermions. In the limit of massless
quarks, the QCD Lagrangian is invariant under $U_L(N_f) \times
U_R(N_f)$ transformations, but the ground state is not. The analysis
of the hadron spectrum and numerical simulations on the lattice
strongly support this assertion \cite{DeTar,Smilref}.
The order parameter of the
spontaneous breaking of chiral symmetry is the chiral quark-antiquark
condensate. In the case of two-color QCD with fundamental
fermions and in the case of any-color QCD with adjoint
fermions the symmetry of the Lagrangian is
enlarged from  $U_L(N_f) \times U_R(N_f)$ to $U(2N_f)$. In the case
of $N_c =2$,  this symmetry is
sometimes referred to as the Pauli-G\"ursey symmetry \cite{pauli-gursey}.
Also in these cases there is strong evidence both from lattice simulations
 \cite{DuKo,DKPR}  and from arguments based on supersymmetry \cite{SUSY} 
that chiral symmetry
is broken by a nonzero vacuum expectation value of the chiral condensate.

The pattern of chiral symmetry breaking is determined by two ideas.
The Vafa-Witten theorem \cite{VafaW} which tells us that vector
symmetries cannot be spontaneously broken, and the idea of maximum
breaking of the axial symmetry \cite{Shifman-three,Peskin}.  For each
of the three classes of theories chiral symmetry is broken
spontaneously according to different patterns
\cite{Peskin,Shifman-three,SmilV}.  The axial U(1)$_A$
subgroup of the global symmetry is broken explicitly by the axial
anomaly in all three cases. The  vector-like $U(1)_B$
symmetry, corresponding to baryon charge conservation, is intact.
The remaining symmetries in QCD with three or more colors 
with fundamental fermions are broken according to $SU_R(N_f)\times SU_L(N_f)
\rightarrow SU_V(N_f)$. For two-color QCD with fundamental fermions
the symmetry is broken according to $SU(2N_f) \rightarrow Sp(2N_f)$,
whereas
for any-color QCD with adjoint fermions the pattern of
symmetry breaking is given by $SU(2N_f) \rightarrow O(2N_f)$.

We can classify these above three cases by the Dyson index, $\beta$,
of the Dirac operator with a value of $\beta =2$, $\beta = 1$ and
$\beta = 4$, respectively. The value of $\beta$ is given by the
number of independent degrees of freedom per matrix element and is
determined by the antiunitary symmetries of the Dirac operator. It is
a concept that originated in Random Matrix Theory \cite{Dyson,SV,V},
and is important for the Cartan classification of symmetric 
spaces \cite{Zirnall}.

In the case of two-color QCD the pseudo-real nature of $SU(2)_{\rm
color}$ \cite{Peskin,Shifman-three,SmilV,Kyoto} can be expressed as
the antiunitary symmetry of the Dirac operator ${\cal D} = \gamma_\nu
D_\nu + m$,
\begin{equation}\label{beta=1} 
[{\cal D}, \tau_2C\gamma_5 K] = 0 
\quad \mbox{ or } \quad
 {\cal D}\tau_2C\gamma_5 = \tau_2C\gamma_5{\cal D}^*,
\hfil \qquad (\beta=1)
\end{equation}
where $\tau_2$ is the color symmetry generator, $C$ is the Dirac
charge conjugation matrix and $K$ is the complex conjugation operator.
Since $(\tau_2 C K)^2 =1$ it is always possible to find a basis in
which the Dirac operator becomes real which gives $\beta =1$.  

The symmetry (\ref{beta=1}) persists even at $\mu\neq0$ \cite{HOV}.
The reality of the Dirac determinant and the feasibility of lattice Monte
Carlo simulations is the consequence of (\ref{beta=1}). This property
also allows us to use QCD inequalities at {\em finite} $\mu$ to show that
condensation can only occur in the scalar diquark channel \cite{KST}.

In the case of QCD with adjoint quarks
the antiunitary symmetry of the Dirac operator is
\begin{equation}\label{beta=4}
[{\cal D}, C\gamma_5K] =0
\quad \mbox{ or } \quad
{\cal D}C\gamma_5  = C\gamma_5{\cal D}^*.
\hfil\qquad (\beta=4)
\end{equation}
Since $(CK)^2 = -1$, it is always possible to find a basis in which the
Dirac operator can be organized into selfdual (pseudoreal) 
quaternions. The value
of the Dyson index is thus $\beta = 4$.

In both above cases, $\beta=1$ and $\beta=4$,
the antiunitary symmetry leads to enlargement
of the global symmetry to $U(2N_f)$ (at $\mu=0$).

There is no antiunitary symmetry in the case of QCD with three or more colors
with fundamental quarks. The Dirac operator is 
a complex matrix, thus $\beta = 2$.

In this paper we shall study two classes of theories:
 $\beta=1$ and $\beta=4$, at finite chemical potential $\mu$.
The main starting point of our analysis is the observation of the
fact that in these theories the lowest lying baryons belong to the
set of Goldstones of the spontaneously broken extended flavor
symmetry $SU(2N_f)$. As a result the dependence on $\mu$ 
can be described in the Chiral Perturbation Theory framework
 \cite{Weinb,GaL,TV}.

We shall construct the effective Lagrangian
governing the low-momentum modes in the theory. These
modes are the Goldstone particles of the spontaneously broken
global symmetries. The symmetry of the theory is largest at
$\mu=0$,  $m=0$, and so is the number of true Goldstone modes. 
A  nonzero chemical potential $\mu$ and/or a bare
quark mass $m$ removes part of the symmetry and some of
the Goldstone modes acquire masses. Our main goal is to find the
functional dependence of the masses of such pseudo-Goldstones
on $\mu$ and $m$. As we proceed we learn about many other properties
of the theory: condensates and vacuum alignment, phase transitions,
etc.

The two cases: two-color fundamental quarks ($\beta=1$) and any-color
adjoint quarks ($\beta=4$) can be analyzed in a similar way. We shall
perform this analysis as follows. At each step we shall
begin with $\beta=1$ case, explaining the concepts and ideas. 
Then we follow it immediately with the same analysis for the
$\beta=4$ case with emphasis on the comparison between the two cases,
which will help understand both cases better.  We shall use similar
notations for the objects which are conceptually the same in both
cases. In many instances we need not rewrite the formulas, only
changing the meaning of the notations suffices. 
All the formulas which use any properties specific to either
$\beta=1$ or $\beta=4$ are explicitly tagged. The formulas
without such tags are general and apply to both cases.  As we shall quickly
see the two cases naturally complement each other and are, in a
certain sense, dual to each other.

\section{Global symmetries at $m=\mu=0$}
\label{sec:globsym}

\subsection{$\beta=1$}

The fermionic part of the QCD Lagrangian 
with  2 fundamental colors is given by
\begin{equation}\label{lagr2c}
{\cal L} = \bar\psi \gamma_\nu D_\nu\psi
=i\left(\begin{array}{c} \psi_L^* \\  \psi_R^* \end{array}\right)^T
\left(\begin{array}{cc}\sigma_\nu D_\nu
 &0\\0&-\sigma^\dagger_\nu D_\nu \end{array}
\right)
\left(\begin{array}{c}\psi_L\\\psi_R\end{array}\right).
\end{equation}
We are working in Euclidean space with hermitian $\gamma$-matrices
and we use spin matrices $\sigma_\nu=(-i,\sigma_k)$. The quark flavor
(as well as color and spin)
indices are suppressed and the sum over $N_f$ flavors is implied.
The symbol $D_\nu$ denotes color covariant derivative, $\partial_\nu +
iA_\nu$, an antihermitian operator, with $A_\nu$ being a matrix 
in color algebra, $A_\nu = A_\nu^a {\tau_a}/2$. As usual $\bar \psi
= \psi^{\dagger} \gamma_0=\psi^{*\,T}\gamma_0$ and in the Euclidean partition
function $\psi$ and $\psi^*$ are independent integration variables.

The fact that the Lagrangian (\ref{lagr2c}) has a higher flavor symmetry
than the apparent $U(N_f)\times U(N_f)$ is related to the
pseudoreality of the two-color Dirac operator.  In particular, the
conjugate field $\tilde \psi_R = \sigma_2\tau_2  \psi_R^*$
transforms similarly (in the same color representation) to
$\psi_L$.
Rewriting the Lagrangian (\ref{lagr2c}) we find
\begin{equation}\label{lagrsu2nf}
{\cal L}
=i\left(\begin{array}{c} \psi_L^* 
 \\ \tilde\psi_R^*\end{array}\right)^T
\left(\begin{array}{cc}\sigma_\nu  D_\nu&0\\0&\sigma_\nu D_\nu \end{array}
\right)
\left(\begin{array}{c}\psi_L\\\tilde \psi_R\end{array}\right)
=i \Psi^\dagger \sigma_\nu D_\nu \Psi,
\end{equation}
where we have used the well-known properties of the Pauli matrices,
$-\sigma_2 \sigma_\nu^\dagger \sigma_2 = \sigma_\nu^T$ and $ 
-\tau_2\tau_k\tau_2 = \tau_k^T$, taken into account anticommutativity
of  Grassman variables, dropped total derivatives,
and introduced the 
spinor of dimension $2N_f$,
\begin{equation}\label{Psi}
\Psi \equiv
\left(\begin{array}{c}\psi_L\\\sigma_2\tau_2 \psi_R^*\end{array}\right)
\equiv\left(\begin{array}{c}\psi_L\\\tilde \psi_R\end{array}\right).
\hfil \qquad (\beta=1)
\end{equation}
In this form the $U(2N_f)$ symmetry becomes manifest. Due to the
axial anomaly the symmetry in the corresponding quantum theory
is only $SU(2N_f)$ (up to discrete symmetries).

\subsection{$\beta=4$}

QCD with  quarks in the adjoint representation of the color group
is described by the Lagrangian
as in (\ref{lagr2c}), but with different notations. The fields
$\psi$ are now transforming according to  the adjoint representation of the
color group. The color covariant derivative is again given by 
$D_\nu = \d_\nu + iA_\nu$, but now $A_\nu$ is given by
the antisymmetric matrix  $(A_\nu)^{bc} = A^a_\nu f_a^{bc}$,
where $f_a^{bc}=f^{abc}$ are the
generators of the adjoint representation, i.e., the structure constants.

The antisymmetric property of the structure constants now replaces
the property of the fundamental generators: $\tau_2\tau_k\tau_2 =
-\tau^T_k \quad \to \quad f_a^{bc} = - f_a^{cb}$.
Using this property
we can again recast the Lagrangian using  spinors of length $2N_f$
and obtain (\ref{lagrsu2nf}), but with the spinors $\Psi$ 
(and $\tilde\psi_R$) defined by
\begin{equation}\label{spinorsadj}
\Psi \equiv
\left(\begin{array}{c}\psi_L\\ \sigma_2 \psi_R^*\end{array}\right)
\equiv\left(\begin{array}{c}\psi_L\\\tilde \psi_R\end{array}\right).
\hfil (\beta=4)
\end{equation}
The only difference from (\ref{Psi}) is the absence of color $\tau_2$
matrix in the definition of $\tilde\psi_R$, and, of course, the fact that
the spinors $\psi$ carry an adjoint, instead of a fundamental, color index.
Similarly, in terms of the spinors (\ref{spinorsadj})
the $SU(2N_f)$ symmetry of the theory becomes manifest.

\section{Spontaneous symmetry breaking and Goldstones}
\label{sec:breaking}

\subsection{$\beta=1$}

Let us now understand transformation properties of $\langle \bar \psi \psi
\rangle$, the order
parameter of the chiral symmetry breaking, 
with respect to the $SU(2N_f)$ symmetry of the theory.
We can rewrite $\bar\psi\psi$ as
\begin{equation}
\bar\psi\psi 
= \left(\begin{array}{c} \psi_L^*\\  \psi_R^* \end{array}\right)^T
\left(\begin{array}{cc}0&1\\1&0\end{array}\right)
\left(\begin{array}{c}\psi_L\\\psi_R\end{array}\right)
=\frac12\Psi^T \sigma_2\tau_2
\left(\begin{array}{cc}0&-1\\1&0\end{array}\right)
\Psi
+ {\rm h.c.} \, .
\hfil (\beta=1)
\label{condens1}
\end{equation}
The Pauli matrices $\sigma_2$ and $\tau_2$
ensure antisymmetrization in spin
and color indices, to produce a spin and color singlet.
We see that the condensate is not invariant under all $SU(2N_f)$
rotations. The subgroup which leaves (\ref{condens1}) invariant is
$Sp(2N_f)$. The Goldstone manifold is therefore given by 
$SU(2N_f)/ Sp(2N_f)$ with $N_f(2N_f-1)-1$ independent degrees of freedom.

Another way of counting the total number of Goldstone modes is
starting from the observation that the condensate (\ref{condens1})
is a product of two fundamental $SU(2N_f)$ flavor representations, 
antisymmetric in flavor indices. Therefore, the condensate belongs
to an antisymmetric tensor representation the dimension of which 
is $N_f(2N_f-1)$.  Condensation can occur in any of these
$N_f(2N_f-1)$ directions; the fluctuations along the
remaining $N_f(2N_f-1)-1$ directions then become Goldstone modes.

The effective theory for the Goldstone modes can be written
in terms of the fluctuations of the orientation of the 
chiral condensate, $\Sigma$, which, according to the previous
paragraph, is an antisymmetric
unimodular ($\det\Sigma=1$) unitary matrix 
(exactly $N_f(2N_f-1)-1$ independent components).
If we denote the equilibrium value of the orientation
of the chiral condensate by $\Sigma_c$,
the Goldstone manifold given by $SU(2N_f)/Sp(2N_f)$
can be parameterized, according to the transformation of $\Sigma$
under $SU(2N_f)$, by
\begin{equation}\label{ssc}
\Sigma = U\Sigma_c U^T,  
\end{equation}
where 
\be
\label{usig}
U = \exp\left(i\Pi\over2F\right)
\mbox{  and  } 
\Pi = \pi_a{X_a\over\sqrt{2N_f}}.
\ee
The fields $\pi_a$ are the Goldstone modes.
The implied sum (over $a$) is over the $N_f(2N_f-1)-1$ generators of the coset 
$SU(2N_f)/Sp(2N_f)$ and in order to simplify the algebra in later sections
we are using the normalization
\begin{equation}
\Tr X_a X_b = 2N_f\delta_{ab}.
\end{equation}

The construction of the Goldstone manifold (\ref{usig}) corresponds to the
classification of the 
$(2N_f)^2-1$ generators of the $SU(2N_f)$ with respect to
a fixed antisymmetric antiunitary matrix, in our case given by $\Sigma_c$,
into $T_k$ and $X_a$ \cite{Peskin}. 
The $T_k$ generators leave $\Sigma_c$ invariant,
\begin{equation}
\exp(i\phi_k T_k)\Sigma_c\exp(i\phi_k T_k)^T = \Sigma_c, 
\mbox{ i.e. } T_k\Sigma_c = - \Sigma_c T_k^T.
\label{peskin}
\end{equation}
By definition, they are the generators of the symplectic group $Sp(2N_f)$.
The remaining  generators, $X_a$,  form the coset $SU(2N_f)/Sp(2N_f)$. 
They obey the relation
\begin{equation}\label{xsigma}
X_a\Sigma_c = \Sigma_c X_a^T, \quad \mbox{ and thus   } \quad 
 U\Sigma_cU^T = U^2\Sigma_c.
\end{equation}

The partition of  generators in generators of the coset, $X_a$, and
generators of the invariant subgroup $Sp(2N_F)$, $T_i$, 
depends on the matrix $\Sigma_c$.
The defining relations (\ref{xsigma}) are left unaltered by rotation
of $\Sigma_c$  according to
\be\label{s-vsv}
\Sigma_c \to  V \Sigma_c V^T,
\ee
where $V$ is an $SU(2N_f)$ matrix, and a simultaneous 
rotation of the generators by
\be\label{x-vxv}
X_a \to VX_a V^\dagger.
\ee
This  means that 
the set of broken generators, $X_a$, changes, if the matrix $\Sigma_c$
is changed.

If we use, as we do below, the following choice for $\Sigma_c$:
\begin{equation}\label{sigmac}
\Sigma_c = \left(\begin{array}{cc}0&-1\\1&0\end{array}\right),
\hfil (\beta=1),
\end{equation}
the 
generators $X_a$ can be written in the block 
representation as
\begin{equation}\label{PiPQ}
\Pi  = \left(\begin{array}{cc}P^T&Q\\Q^\dagger&P\end{array}\right),
\end{equation}
with $N_f\times N_f$ matrices $P$ and $Q$ such that $\Tr P = 0$,
$P^\dagger = P$, and $Q^T=-Q$. It is then easy to see that
the number of independent components in $P$ and $Q$ are,
\begin{equation}\label{NPQ}
N_P = N_f^2-1 \quad\mbox{ and }\quad N_Q = N_f(N_f-1).
\hfil \qquad(\beta=1)
\end{equation}

\subsection{$\beta=4$}

What are the transformation properties of the chiral condensate
with respect to the $SU(2N_f)$ symmetry in the $\beta=4$ case?
We can rewrite $\bar\psi\psi$ as
\begin{equation}\label{condadj}
\bar\psi\psi 
= \left(\begin{array}{c} \psi_L^* \\ \psi_R^*\end{array}\right)^T
\left(\begin{array}{cc}0&1\\1&0\end{array}\right)
\left(\begin{array}{c}\psi_L\\\psi_R\end{array}\right)
=\frac12\Psi^T \sigma_2
\left(\begin{array}{cc}0&-1\\-1&0\end{array}\right)
\Psi
+ {\rm h.c.}\, .
\hfil \qquad(\beta=4)
\end{equation}
The Pauli matrix $\sigma_2$ ensures the antisymmetrization with respect
to spin indices, to produce a spin singlet. There is no
antisymmetrization in color (unlike in the case of fundamental colors
$\beta=1$) --- the color singlet is a symmetric product of two
adjoint representations. The Pauli principle now demands that the
$SU(2N_f)$ flavor indices must be symmetric, and they are,
as is evident in (\ref{condadj}). Therefore the condensate belongs
to the symmetric (as opposed to antisymmetric for $\beta=1$) 
second rank tensor
representation of $SU(2N_f)$, which has dimension $N_f(2N_f+1)$.
The condensation can occur in any of the $N_f(2N_f+1)$ directions.
The fluctuations in the remaining $N_f(2N_f+1)-1$ directions
become Goldstone bosons. Alternatively, 
since the chiral condensate (\ref{condadj})
is invariant under $SO(2N_f)$, the Goldstone manifold is
$SU(2N_f)/SO(2N_f)$, which gives us the same number of Goldstone modes.

The Goldstone manifold $SU(2N_f)/SO(2N_f)$ should now be parameterized
by {\em symmetric} unimodular unitary matrices $\Sigma$.
The Goldstone fields are introduced in the same way as 
in (\ref{ssc}) and (\ref{usig}), i.e., $\Sigma=U\Sigma_cU^T$,
where $\Sigma_c $ is now also a {\em symmetric} unimodular 
unitary matrix and the implied sum in (\ref{usig}) is over the generators 
$X_a$ of the coset $SU(2N_f)/SO(2N_f)$. 

The classification of the generators is also similar to $\beta=1$
case. The generators of the coset, $X_a$ obey the commutation
relation (\ref{xsigma}) with a given symmetric unitary matrix, also 
denoted by $\Sigma_c$,
whereas the $T_i$ generators leave $\Sigma_c$ invariant as in
(\ref{peskin}). 
The $T_i$ are now the generators
of  an $SO(2N_f)$ subgroup (as opposed to $Sp(2N_f)$ in the
$\beta=1$ case). 
The remaining generators $X_a$ form a coset
$SU(2N_f)/SO(2N_f)$, which is the Goldstone manifold in this case.

In this case the standard choice for the matrix $\Sigma_c$ is
\begin{equation}\label{sigmacadj}
\Sigma_c = \mat 0 & 1 \\ 1 & 0 \emat.
\hfil (\beta = 4)
\end{equation}
With this choice, the matrix $\Pi =\pi_a X_a/\sqrt{2N_f}$ in the coset can 
be split into 4 blocks of size
$N_f\times N_f$ as in (\ref{PiPQ}). The matrix $P$ is again hermitian
and traceless, but the matrix $Q$ is now {\em symmetric}. 
Therefore, the counting
of the independent degrees of freedom in the $\beta=4$ case is
\begin{equation}\label{NPQadj}
N_P = N_f^2-1 \quad\mbox{ and }\quad N_Q = N_f(N_f+1). \hfil (\beta=4)
\end{equation}

In both cases, $\beta=1$ and $\beta=4$,
the kinetic term of the effective Lagrangian
describing the Goldstone modes should be invariant
under the global $SU(2N_f)$ group and under Lorentz transformation.
The corresponding $SU(2N_f)$ nonlinear sigma-model is given by
\begin{equation}\label{Lnomass}
{\cal L}_{\rm eff} = {F^2\over2} \Tr\d_\nu\Sigma\d_\nu\Sigma^\dagger,
\end{equation}
where $F$ is the pion decay constant.

\section{Bare quark mass $m$}
\label{sec:baremass}

\subsection{$\beta=1$}
\label{sec:baremass1}

If the bare quark mass $m$ (in this paper, the same for all quarks)
is not zero, an explicit $SU(2N_f)$ breaking term in the effective Lagrangian
(\ref{Lnomass}) appears. 
To determine its form we first rewrite the bare mass term
using $SU(2N_f)$ notations,
\begin{equation}\label{baremass}
m\bar\psi\psi = \frac 12 m\Psi^T \sigma_2 \tau_2
\left(\begin{array}{cc}0&-1\\1&0\end{array}\right)
\Psi +{\rm h.c.}= -\frac 12\Psi^T \sigma_2\tau_2 M \Psi +{\rm h.c.},  
\hfil (\beta=1)
\end{equation}
where the mass matrix $M$ is given by 
\begin{equation}
M= m \hat M \quad {\rm and} \quad \hat M = \mat 0 & 1 \\ -1 & 0 \emat.
\hfil (\beta=1)
\end{equation}
We see explicitly that the bare mass term is 
only invariant under an $Sp(2N_f)$ subgroup of $SU(2N_f)$.
The full $SU(2N_f)$
invariance can be restored if $M$ is also transformed
together with $\Psi$ according to
\begin{equation}
\Psi \to V\Psi \quad \mbox{ and } \quad M\to V^* M V^\dagger.
\end{equation}
This extended symmetry must be also manifest in the
effective theory,
\begin{equation}
\Sigma\to V\Sigma V^T \quad \mbox{ and }\quad  M\to V^* M V^\dagger.
\end{equation}
The lowest order term induced by the quark mass must therefore
have the form
\begin{equation}\label{q.mass}
{\cal L}_{\rm q.mass} = -G {\rm Re }\Tr (M\Sigma)
= -m G {\rm Re }\Tr (\hat M \Sigma).
\end{equation}
One can view ${\rm Re }\Tr (\hat M\Sigma)$ as a generalized cosine of the angle
between unitary matrices $\Sigma$ and $\hat M^\dagger$. It is
maximal when $\Sigma$ is aligned with $\hat M^\dagger$.
Therefore the direction of $\Sigma$
minimizing (\ref{q.mass}) is given by
\begin{equation}\label{sigmacM}
\Sigma_c = \hat M^\dagger ,
\end{equation}
which leads us to our choice of $\Sigma_c$ (\ref{sigmac}).
The mass term comes with a phenomenological coefficient, which 
we denote by $G$. It is given by the derivative
of the vacuum energy with respect to $m$ and is, therefore, proportional
to the chiral condensate in the chiral limit $m\to0$ at $\mu=0$
(see Section \ref{sec:condensates}),
\begin{equation}
G=\frac{\langle\bar\psi\psi\rangle_0}{2N_f}.
\end{equation}
The resulting Lagrangian with the mass term, 
\begin{equation}\label{Lmu=0}
{\cal L}_{\rm eff} = {F^2\over2} \Tr\d_\nu\Sigma\d_\nu\Sigma^\dagger -
m G {\rm Re }\Tr (\hat M \Sigma),
\end{equation}
is the familiar Chiral Perturbation Theory Lagrangian at lowest order in the
momentum expansion \cite{Weinb,GaL}.
Expanded to second order in the pion fields according to (\ref{ssc}),
(\ref{usig}), it yields a spectrum with $N_f(2N_f-1)-1$ degenerate
(pseudo-)Goldstones with masses given by the usual Gell-Mann$-$Oakes$-$Renner
relation
\begin{equation}\label{GOR}
m_\pi^2 = {m G \over F^2}.
\end{equation}
We can use this relation to trade $G$
for another parameter, $m_\pi$, and write
\begin{equation}\label{leffmpi}
{\cal L}_{\rm eff} = {F^2\over2} \left[
\Tr\d_\nu\Sigma\d_\nu\Sigma^\dagger 
-
2m_\pi^2 {\rm  Re}\,\Tr (\hat M \Sigma).
\right].
\end{equation}

The symmetry of the theory is reduced from $SU(2N_f)$ to $Sp(2N_f)$
by the mass term. Since the chiral condensate does not break any
more symmetries in this case we do not have any true Goldstones
when $m\ne0$.

\subsection{$\beta=4$}

To determine the form of the term in the effective Lagrangian
induced by a small bare quark mass $m$ we rewrite the quark mass term as
\begin{equation}\label{baremassadj}
m\bar\psi\psi = \frac 12 m\Psi^T \sigma_2
\left(\begin{array}{cc}0&-1\\-1&0\end{array}\right)
\Psi  + {\rm h.c.}= -\frac 12 \Psi^T \sigma_2 M \Psi + {\rm h.c.}\,,
\hfil (\beta=4)
\end{equation}
where we have used the spinors of length $N_f$ introduced in section 
\ref{sec:globsym}.
In this case the mass matrix is given by
\begin{equation}\label{mmm4}
M = m \hat M \quad {\rm with } \quad \hat M = \mat 0 & 1 \\ 1 &0 \emat.
\hfil(\beta=4)
\end{equation}
Using the same arguments as in section~\ref{sec:baremass1}
we find  that the mass term in the effective Lagrangian
as dictated  by the extended flavor symmetry 
is given again by (\ref{q.mass}), with $\Sigma$ now being a symmetric
matrix and $M$ given by (\ref{mmm4}).
Similarly, this term is minimized when $\Sigma=\hat M^\dagger$. 
With $\hat M$ now taken from (\ref{mmm4}) we arrive at our choice of
$\Sigma_c$ (\ref{sigmacadj}).

The symmetry of the theory is reduced from $SU(2N_f)$ to $SO(2N_f)$ by the mass
term. Since the chiral  condensate does not break any more symmetries 
in this case
we do not have any truly massless Goldstones when $m\ne0$.  The masses of all
$N_f(2N_f+1)$ \mbox{(pseudo-)Goldstones} are equal and are given by the
Gell-Mann$-$Oakes$-$Renner relation (\ref{GOR}).

\section{Chemical potential $\mu$}
\label{sec:mu}

In this section we review the introduction of the chemical potential
in the effective partition function following the approach of 
\cite{KST}. This approach relies only on the $SU(2N_f)$ symmetry 
of the theory and the resulting
$\mu$-dependent terms are the same for both cases,
$\beta=1$ and $\beta=4$.

\subsection{Global symmetries and $\mu$}

At nonzero chemical potential the microscopic Lagrangian is given by
\begin{equation}
{\cal L} 
=\bar\psi \gamma_\nu D_\nu\psi - \mu\bar\psi\gamma_0\psi +m\bar \psi \psi.
\end{equation}
As was the case for the mass term, we can also rewrite 
the baryon charge density in terms of the $SU(2N_f)$ spinors.
\begin{eqnarray}\label{pb0p}
&&\bar\psi\gamma_0\psi =
\left(\begin{array}{c}\psi_L^*\\ \psi_R^*\end{array}\right)^T
\left(\begin{array}{cc}1&0\\0&1\end{array}\right)
\left(\begin{array}{c}\psi_L\\\psi_R\end{array}\right)
= \Psi^\dagger\left(\begin{array}{cc}1&0\\0&-1\end{array}\right)\Psi
\equiv \Psi^\dagger B \Psi;
\nonumber\\
&& B \equiv \mat +1&0\\0&-1\emat.
\end{eqnarray}
The physical meaning of $+1$ and $-1$ in the baryon charge matrix $B$ 
is simple: they are  the baryon charges of the quarks $\psi_L$ and 
conjugate quarks $\tilde \psi_R$.
We see that this term is not a singlet under $SU(2N_f)$. It transforms
in the adjoint representation of this group, in other words,
the baryon charge is one of the $(2N_f)^2 - 1$ generators of this group.
In terms of the spinors of length $2N_f$, the microscopic Lagrangian
is thus given by
\be
{\cal L} = i \Psi^\dagger\sigma_\nu (D_\nu - \mu B_\nu) \Psi
-\left [ \frac 12 \Psi^T \left \{ \begin{array}{c} 
\sigma_2 \tau_2 \\ \sigma_2 \end{array} \right \} M \Psi + {\rm h.c.}
 \right ],
\label{LmicroB}
\ee
where the upper branch corresponds to $\beta = 1$ and the lower
branch to $\beta = 4$. For reasons that will become clear in the next
subsection, we have introduced the four-vector $B_\nu=(B,{\bf 0})$.

As in the case of the quark mass term, the chemical potential term,
$\mu\Psi^\dagger B \Psi$ violates the $SU(2N_f)$ symmetry. Similarly,
we can maintain this symmetry by accompanying the rotation of $\Psi$
by a corresponding rotation of $B$,
\begin{equation}\label{sympsib}
\Psi\to V\Psi \mbox{ and } B\to V B V^\dagger.
\end{equation}
Such an extended symmetry must be manifest in the effective Lagrangian,
\begin{equation}
\Sigma \to V\Sigma V^T \mbox{ and } B\to V B V^\dagger.
\end{equation}
This restricts the lowest order nonderivative term in $\mu$ 
to a linear combination of
\begin{equation}\label{sbsb}
\mu^2 \Tr (\Sigma B^T \Sigma^\dagger B)
\qquad\mbox{ and }\qquad
\mu^2 \Tr ( B  B)
\end{equation}
with arbitrary coefficients. Only the first of these terms contains
a dependence on the Goldstone fields.%
\footnote{Such  type of symmetry breaking terms also occur
in the context of the non-hermitian Random Matrix Theory
\cite{St96,Fyodorov,Efetov}.}

At $m=0$ the chemical potential breaks the global 
symmetry of the theory from $SU(2N_f)$ down to the usual 
(as in three-color QCD) 
$SU(N_f)_L\times SU(N_f)_R\times U(1)_B$.
If $m\ne 0$ also, the symmetry of the theory is only 
$SU(N_f)_V\times U(1)_B$.

\subsection{Local symmetry and the coefficient of the $\mu^2$ term}
\label{sec:muloc}

The coefficients of the terms (\ref{sbsb}) 
in the effective Lagrangian can be related
to baryon number susceptibility, i.e., the second derivative of the
vacuum energy with respect to $\mu$. However, unlike $G$, i.e. the
chiral condensate $\langle\bar\psi\psi\rangle_0$, these parameters
are not independent. They are related to the pion decay
constant $F$ by virtue of a local symmetry \cite{KST}.

As was observed in \cite{KST}, one can further extend the symmetry
(\ref{sympsib}) of the microscopic Lagrangian (\ref{LmicroB}) 
 to include {\em local} $SU(2N_f)$ flavor transformations
\begin{equation}
\Psi\to V\Psi 
\quad\mbox{ and }\quad 
B_\nu\to V B_\nu V^\dagger - \frac1\mu V\partial_\nu V^\dagger.
\end{equation}
We can also recover the Lorentz (Euclidean) symmetry by 
transforming $B_\nu$ as a four-vector.

To make such an extended local symmetry (and also Lorentz symmetry) 
manifest in the effective Lagrangian
(\ref{leffmpi}) we must replace the normal derivative by a flavor
covariant derivative,
\begin{eqnarray}
&&\nabla_\nu \Sigma = \partial_\nu \Sigma - \mu(B_\nu\Sigma + \Sigma B_\nu^T);
\nonumber\\
&&\nabla_\nu \Sigma^{\dagger} = \partial_\nu \Sigma^{\dagger} +
\mu(\Sigma^{\dagger}B_\nu + B_\nu^T\Sigma^{\dagger}).
\end{eqnarray}
Thus we arrive at the effective Lagrangian of lowest order in the momentum
expansion 
\begin{eqnarray}\label{L}
{\cal L}_{\rm eff} &=& {F^2\over2} \left[
{\rm Tr} \nabla_\nu\Sigma \nabla_\nu\Sigma^\dagger
-
2m_\pi^2 {\rm \Re} {\rm Tr} (\hat M\Sigma)
\right]
\nonumber\\&& 
= {F^2\over2} {\rm Tr} \partial_\nu\Sigma \partial_\nu\Sigma^\dagger
+ 2\mu F^2 {\rm Tr} B\Sigma^\dagger\partial_0\Sigma
\nonumber\\&& 
\hskip 1em -F^2
\mu^2 {\rm Tr} \left(\Sigma B^T\Sigma^\dagger B + B B\right)
-F^2 m_\pi^2 {\rm Re} {\rm Tr}\left(\hat M\Sigma\right).
\end{eqnarray} 
It is important to note that the dependence on $\mu$ comes with no
additional parameters. It is completely fixed, by the
local symmetry, in terms of an already existing parameter
$F$.

\section{Vacuum alignment}
\label{sec:vacuum}

The static part of the Lagrangian (\ref{L}) determines
the vacuum alignment of the field $\Sigma$ 
as well as the masses of the
excitations. This part of the Lagrangian has the form
\begin{eqnarray}\label{Lstat}
{\cal L}_{\rm st}(\Sigma)&=&-F^2
\mu^2\Tr\left(\Sigma B^T \Sigma^\dagger B +BB\right)
- 
F^2m_\pi^2\Re\Tr (\hat M\Sigma)
\nonumber\\
&=&{F^2{m_{\pi}^2}\over2}\left[
-{x^2\over 2} \Tr\left(\Sigma B^T \Sigma^\dagger B +BB\right)
- 2\Re\Tr (\hat M\Sigma)
\right],
\end{eqnarray}
where we introduced $x=2\mu/m_{\pi}$. The $\mu^2$ and $m_\pi^2$
 terms in (\ref{Lstat})
compete for the direction of the condensation
which we denote by $\overline\Sigma$. For 
$x=0$ the orientation of $\overline\Sigma$ is determined
by the mass matrix $\overline\Sigma=\hat M^\dagger$. This
value we
denoted by $\Sigma_c$. This is the orientation of the usual chiral
condensate which carries no baryon charge.
When $x=\infty$ \cite{KST} the static Lagrangian is minimized on a manifold
in the space of $\Sigma$, from which we choose the following value
and denote it by $\Sigma_d$, 
\begin{equation}\label{sigmad}
\Sigma_d = \left(\begin{array}{cc}iI&0\\0&iI\end{array}\right) 
,
\hfil (\beta=1)
\end{equation}
where $I$ is an antisymmetric $N_f\times N_f$ matrix,
which, written in $(N_f/2)\times(N_f/2)$ blocks, has the form
\begin{equation}\label{I}
I = \left(\begin{array}{cc}0&-1\\1&0\end{array}\right).
\end{equation}
This minimum (even for $m \ne 0$, unlike $\Sigma_c$) has a degeneracy, which 
for $N_f =2$, is simply a rotation by the 
generator $B$. The condensate $\Sigma_d$
breaks spontaneously the baryon number symmetry. The degeneracy leads to
the appearance of a massless  Goldstone boson. For $N_f>2$ this
condensate breaks more than just the $U(1)_B$ symmetry. For 
$m=0$ it also breaks 
$SU(N_f)_L\times SU(N_f)_R$ down to $Sp(N_f)_L\times Sp(N_f)_R$ given
by the rotations leaving the block matrix $I$
in (\ref{sigmad}) invariant \cite{KST}. When $m\ne 0$ 
it breaks the $SU(N_f)_V$ symmetry down to $Sp(N_f)_V$.%
\footnote{Similarly to the alignment of $\Sigma_c$ to the bare quark 
mass matrix $M^\dagger$, the direction of $\Sigma_d$
is determined by an external diquark source, $J$, which, 
as we shall see in section \ref{sec:dqsource}, breaks the degeneracy.
When $J$ is zero, we can choose any orientation of $\Sigma_d$
within the manifold of minima, and the results will not
change due to the symmetry relating all such minima.}

The situation is similar in the $\beta=4$ case. The only difference is
that now the matrix $\Sigma$ is a {\em symmetric} $SU(2N_f)$ matrix.
It rotates between the value of $\Sigma_c$ given by
(\ref{sigmacadj}) and $\Sigma_d$, which we chose as
\begin{equation}
\label{sigmadadj}
\Sigma_d = \left(\begin{array}{cc}i&0\\0&i\end{array}\right) 
.
\hfil (\beta=4)
\end{equation}
The discussion of the preceding paragraph carries over to the $\beta=4$
case with the already familiar substitution of $Sp(N_f)$ by $SO(N_f)$.
The condensate $\Sigma_d$ breaks $SU(N_f)_{L,R}$ flavor
symmetries down to $SO(N_f)_{L,R}$, in addition to breaking $U(1)_B$.

At intermediate values of $x$ the orientation of the 
condensate $\overline\Sigma$ 
rotates, as a function of
$x$, from $\Sigma_c$ to $\Sigma_d$. We shall prove that it can  always
be written as the linear combination
\begin{equation}\label{osigma}
\overline\Sigma = \Sigma_\alpha \equiv
\Sigma_c \cos\alpha + \Sigma_d \sin\alpha,
\end{equation}
where $\alpha=0$ for $x=0$ and $\pi/2$ for $x=\infty$.

The angle $\alpha$ is a function of $x$. It is determined
by substituting the value of $\Sigma$ given by (\ref{osigma}) into
(\ref{Lstat}). This results in the static Lagrangian
\begin{equation}\label{Leffalpha}
{\cal L}_{\rm st}(\Sigma_\alpha) = 
F^2{m_{\pi}^2}N_f
\left[{x^2\over 2} (\cos2\alpha - 1) - 2\cos\alpha\right] .
\end{equation}
Minimizing with respect to
$\alpha$  we find
\begin{eqnarray}\label{alpha}
\alpha = 0, &&\mbox{ when } x < 1;\nonumber\\
\cos\alpha = {1\over x^2}, &&\mbox{ when } x>1.
\end{eqnarray}
Note that $\alpha=0$ is always an extremum of (\ref{Leffalpha})
but it becomes a maximum for $x>1$.

We find that the condensate is a {\em non-analytic} function 
of $x=2\mu/m_\pi$. There is a 
{\em second} order phase transition as a function of $\mu$ at $x=1$. 
For small $\mu<m_{\pi}/2$ the vacuum does not change. At
$\mu=m_{\pi}/2 $ a transition occurs and the direction of the
condensate 
starts rotating from
that of $\Sigma_c$ to that of $\Sigma_d$.  A nonzero value of the projection
on $\Sigma_d$, proportional to $\sin\alpha$, means that the vacuum
breaks the baryon number symmetry spontaneously. 
This happens at the value of $\mu$ equal to
1/2 of the mass of the lightest baryon in the system -- the
diquark, or, the baryonic pion.

\section{Global minimum}
\label{sec:globalminimum}

In this section we show that the minimum of the static potential is given
by  $\Sigma_\alpha$ defined in 
(\ref{osigma}). The argument in this section shows that it 
is an absolute, or global,
minimum of ${\cal L}_{\rm st}(\Sigma)$. In the next section we expand
the Goldstone fields to second order about this minimum.

\subsection{$\beta=1$}

We decompose the antisymmetric unitary
matrix $\Sigma$ into 4 blocks of size $N_f\times N_f$,
\begin{equation}
\Sigma = \mat A &  -C\\ C^T & B \emat.
\hfil (\beta=1)
\label{sym1}
\end{equation}
The antisymmetric matrices $A$ and $B$ satisfy the following unitarity 
constraints:
\begin{equation}\label{ABC}
AA^\dagger + CC^\dagger = 1,\qquad
BB^\dagger + (C^\dagger C)^T = 1,\qquad
AC^* = CB^\dagger.
\hfil (\beta=1)
\end{equation}
Taking this into account we can express ${\cal L}_{\rm st}$
entirely in terms of the matrix $C$,
\be
{\cal L}_{\rm st}(\Sigma) = 
{F^2{m_{\pi}^2}} \left [
x^2 {\rm Tr } \, \left(C-\frac 1{x^2}\right)
\left(C^\dagger-\frac 1{x^2}\right)
-N_f\left({x^2}+\frac 1{x^2}\right) \right ]
.
\label{cpotential}
\ee

Ignore, for the moment, the constraints on the matrix elements
of $C$. 
The trace in (\ref{cpotential}) can be viewed
as the distance in the $2N_f^2$ dimensional space of real and 
imaginary parts of the matrix elements of $C$ from the
point given by diagonal matrix $1/x^2$. For $x>1$ the absolute minimum
is achieved when $C=1/x^2$. The matrices $A$ and $B$ can be chosen,
for example, as $A=B=iI\sqrt{1-1/x^4}$, to satisfy all the constraints
(\ref{ABC}). If we define $\cos\alpha=1/x^2$, we observe that the resulting 
matrix in (\ref{sym1}) is given by $\Sigma_\alpha$ (\ref{osigma}).

When $x<1$, we notice that unitarity
constraints (\ref{ABC}) demand that $\Tr CC^\dagger\le 1$.
This means that we have to consider the points in the space
of $C$ only within the unit hypersphere around $C=0$. It is easy
to see that the minimum distance within this sphere is
at the surface point closest to $1/x^2$, i.e. at $C=1$.
The constraints (\ref{ABC}) can be satisfied only by $A=B=0$
and the resulting matrix in (\ref{sym1}) is $\Sigma_c$, the minimum
of the static potential at $\mu = 0$.

\subsection{$\beta=4$}

The same analysis applies in the $\beta=4$ case.
Since $\Sigma$ is now a symmetric matrix its
block decomposition is given by
\begin{equation}
\Sigma = \mat A &  +C\\ C^T & B \emat,
\label{sym4}
\hfil (\beta=4)
\end{equation}
with symmetric matrices $A$ and $B$ and with unitarity constraints
\begin{equation}\label{ABCadj}
AA^\dagger + CC^\dagger = 1,\quad
BB^\dagger + (C^\dagger C)^T = 1,\qquad
AC^* = -CB^\dagger.
\hfil (\beta=4)
\end{equation}
The static part of the Lagrangian is the same as in
(\ref{cpotential}). The minimum is given by 
$C = 1$ for $x < 1$ and $ C= 1/x^2$ for $x > 1$. The matrices
$A$ and $B$ are not determined uniquely by the unitarity
constraints. One can check that  
$A= B = i\sqrt{1 -1/x^4}$ (as in (\ref{osigma})) 
satisfy these constraints.
\footnote{
As in the
case of $\beta = 1$, the case of odd number of flavors has to be
investigated separately.
Let us consider $N_f =1$. Then a symmetric
unitary matrix has two degrees of freedom and can be parameterized as
\be
\Sigma = \mat i \sin \theta e^{i\phi} & \cos \theta \\
               \cos \theta &i \sin \theta e^{-i\phi} \emat.
\nonumber
\ee
Notice that the determinant of $\Sigma$ has to be equal to the determinant
of the mass matrix $\hat M$ (\ref{mmm4}).
We thus have that the matrix $C$ in (\ref{sym4}) is just a number
parameterized according to
$C =  \cos \theta$,
leading to the effective potential (up to constants)
\be
x^2{\rm Tr} \, \left(C^\dagger - \frac 1{x^2}\right)
\left( C - \frac{1}{x^2}\right)
-x^2 -\frac 1{x^2} =x^2 \cos^2 \theta-2 \cos \theta-x^2.
\nonumber
\ee
The minimum is at $\theta=0$ for $x<1$ and at $\cos \theta=1/x^2$ for
$x>1$. In the diquark condensation
phase ($x>1$), we find one massless excitation
(the $\phi$-mode), a nonzero 
baryon density and nonzero chiral and diquark condensates.}

\section{Curvatures at the minimum}
\label{sec:curv}

We have established that the global minimum of the
effective Lagrangian is achieved at the value of the field $\Sigma$
given by (\ref{osigma}) and (\ref{alpha}). The orientation of the condensate
rotates as a function of $x=2\mu/m_\pi$ (nonanalytic at $x=1$).
In this section we shall expand the effective Lagrangian
up to the second order in the fluctuations of $\Sigma$ 
which will help us to
determine the (pseudo-)Goldstone masses and their dependence on $x$.

\subsection{Normal phase: $\mu<m_{\pi}/2$}
\label{sec:normal}

When $x<1$ the vacuum orientation of $\Sigma$ does not depend
on $x$ and is given by $\overline\Sigma = \Sigma_c$.
Expanding $\Sigma$ around $\Sigma_c$ using the Goldstone fields
defined in (\ref{usig}) according to
\be\label{sigmaexp}
\Sigma = U\Sigma_cU^T = U^2\Sigma_c =
\left(1 + {i\Pi\over F} - {\Pi^2\over 2F^2} + \ldots\right)\Sigma_c,
\ee
we find
\begin{equation}
{\cal L}_{\rm st}(\Sigma) = {\cal L}_{\rm st}(\Sigma_c)+
{{m_{\pi}^2}\over2} \left[
{x^2\over4}\Tr[B,\Pi]^2 + \Tr \Pi^2
\right] + \ldots,
\end{equation}
where ellipsis denotes terms of higher powers of $\Pi$.
The commutator in this Lagrangian can be written as
\be
[B,\Pi] = b\Pi,
\ee
with $b$ the baryon charge of the 
pseudo-Goldstone modes. Since all our pseudo-Goldstone modes are  
quark-antiquark or (anti-)diquark
states the values of $b$ are  $b=0,\pm2$.
The curvature of the 
$P$, $Q$ and $Q^\dagger$ (\ref{PiPQ}) modes,
thus depends on $\mu$ through the baryon charge
and is given by $m_{\pi}^2-(b\mu)^2$.
 For example, in the $\beta=1$ case of $N_f=2$,
there are 3 pseudoscalar mesons, 1 diquark and 1 antidiquark.
Using the block decomposition of the
generators (\ref{PiPQ}) we find
\begin{equation}
{\cal L}_{\rm st}(\Sigma) = {\cal L}_{\rm st}(\Sigma_c)+
{{m_{\pi}^2}} \Tr \left[
P^2 + (1-x^2) QQ^\dagger
\right] + \ldots.
\end{equation}

Two comments are in order here.
First, note that, the curvature (and the mass) of the
meson modes $P$ does not depend on $\mu$ in the normal phase.
Second, at $x=1$
the curvature of the diquark modes $Q$ vanish, which signals
a phase transition and the onset of diquark condensation.

\subsection{Diquark condensation phase: $\mu>m_{\pi}/2$.}

In this phase the condensate begins to rotate according to
(\ref{osigma}), (\ref{alpha}). This rotation can be also written
as
\begin{equation}\label{vsv}
\Sigma_\alpha = V_\alpha \Sigma_c V_\alpha^T = V_\alpha^2 \Sigma_c, 
\mbox{ where } 
V_\alpha^2 = e^{i\alpha X_2},
\end{equation}
and $X_2$ is the generator that 
rotates $\Sigma_c$ into $\Sigma_d$. Comparing (\ref{vsv})  
(\ref{osigma}) we find
\begin{equation}\label{sx2s}
\Sigma_d = iX_2\Sigma_c.
\end{equation}
This generator belongs to the set of broken generators
with respect to $\Sigma_c$ (as well as with respect to $\Sigma_d$)
since it satisfies (\ref{xsigma}).

We could parameterize fluctuations of $\Sigma$ around the vacuum
value given by $\Sigma_\alpha$ as
\begin{equation}
\Sigma = U_\alpha \Sigma_\alpha U_\alpha^T,
\end{equation}
where $U_\alpha$ are unitary matrices generated by rotated
(in the sense of (\ref{x-vxv}))
generators, $V_\alpha X_aV_\alpha^\dagger$, instead of $X_a$
(\ref{ssc}), (\ref{usig}). 
However, as we
shall find, the meson mass matrix is diagonal in the basis given by
the  parametrization
\begin{equation}\label{vusuv}
\Sigma = V_\alpha U \Sigma_c U^T V_\alpha^T,
\end{equation}
where we have rotated the coset generators $X_a$ back to their $\alpha=0$
values using (\ref{s-vsv}), (\ref{x-vxv}) and (\ref{vsv}).

Before expanding the static Lagrangian, 
we substitute (\ref{vusuv}) into (\ref{Lstat})
to obtain
\begin{eqnarray}\label{Lstatalpha}
{\cal L}_{\rm st}(\Sigma)=
{F^2{m_{\pi}^2}\over2}&&\hskip -1em\left[
-{x^2\over 2} \Tr \left(
U^2\Sigma_c \left({V_\alpha}^\dagger BV_\alpha\right)^T
\Sigma_c^\dagger \left(U^\dagger\right)^2 
\left({V_\alpha}^\dagger B V_\alpha\right)
+ BB\right)
\right.\nonumber\\&&\left.
- 2\Re\Tr 
\left(\left(V_{\alpha}^T \hat M {V_{\alpha}}\right) U^2\Sigma_c\right)
\right].
\end{eqnarray}
The rotated values of $B$ and $\Sigma_c$ (by angle $-\alpha$
in the sense of (\ref{vsv}) ) can be expressed as
\begin{eqnarray}\label{bsrotated}
&&V_\alpha^\dagger B V_\alpha      
=V_{-\alpha} B V_{-\alpha}^\dagger
= B \cos\alpha + X_1 \sin\alpha;
\nonumber\\
&&V_{\alpha}^T \hat M V_{\alpha} 
= (V_{-\alpha} \hat M^\dagger V_{-\alpha}^T)^\dagger 
= \Sigma_{-\alpha}^\dagger
= \Sigma_c^\dagger\cos\alpha - \Sigma_d^\dagger\sin\alpha,
\end{eqnarray}
where we have introduced $X_1$ by
\begin{equation}\label{x1}
X_1 = iBX_2.
\end{equation}
This generator, similarly to $X_2$, belongs to the coset of broken
generators with respect to $\Sigma_c$ (\ref{xsigma}). It is
the generator into which $B$ is rotated while $\Sigma_c$
is rotated into $\Sigma_d$.

Substituting (\ref{bsrotated}) into (\ref{Lstatalpha}) and
expanding to second order in $\Pi$ (\ref{sigmaexp}) we find
\begin{eqnarray}\label{L_exp}
{\cal L}_{\rm st}(\Sigma)&=&{\cal L}_{\rm st}(\Sigma_\alpha)
+ {Fm_\pi^2}\left[-{x^2\over2}\sin2\alpha+\sin\alpha\right]
\Tr\left(X_2\Pi\right)
\nonumber\\&&
+
{{m_{\pi}^2}\over2} \left[
{x^2\over4}\left(\Tr[B,\Pi]^2\cos^2\alpha
-\Tr[X_1,\Pi]^2\sin^2\alpha \right)
+ \Tr \Pi^2\cos\alpha
\right] + \ldots.
\end{eqnarray}
As should be expected, the linear term vanishes due to (\ref{alpha})
and we shall concentrate now on the quadratic term.

\subsubsection{$\beta=1$}

At this point we need the explicit form of $X_1$ for $\beta=1$.
According to (\ref{sx2s}) we have
\begin{equation}\label{x2}
X_2 = -i\Sigma_d\Sigma_c^\dagger = \mat 0&I\\-I&0 \emat.
\hfil (\beta=1)
\end{equation}
For $X_1$ we therefore find 
\begin{equation}
X_1 = iBX_2 = \mat 0&iI\\iI&0 \emat.
\hfil \qquad(\beta=1)
\end{equation}

We then use the block decomposition of $\Pi$ (\ref{PiPQ})
and find
\begin{eqnarray}
{\cal L}_{\rm st}(\Sigma)={\cal L}_{\rm st}(\Sigma_\alpha)+
{{m_{\pi}^2}}  \Tr \left[
-x^2 
\left( 
QQ^\dagger\cos^2\alpha 
- (P_S^2 + Q_RQ_R^\dagger)\sin^2\alpha
\right)
\right.\nonumber\\\left.
+ (P^2 + QQ^\dagger)\cos\alpha
\right] + \ldots.
&&\hfil (\beta=1)
\end{eqnarray}
We have introduced the following projections of $P$ and $Q$:
\begin{eqnarray}\label{PQproj}
P_S = \frac12(P + IP^TI) 
\quad\mbox{ and }\quad 
P_A = \frac12(P - IP^TI);\nonumber\\
Q_R = \frac12(Q + IQ^\dagger I )
\quad\mbox{ and }\quad 
Q_I = \frac12(Q - IQ^\dagger I). 
&&\hfil(\beta=1)
\end{eqnarray}
These projections are orthogonal, $\Tr P_SP_A = {\rm Re}\,\Tr
Q_RQ_I^\dagger=0$.  
Using this fact and the relation $\cos\alpha=1/x^2$
we obtain
\begin{eqnarray}\label{LeffPQ}
{\cal L}_{\rm st}(\Sigma)={\cal L}_{\rm st}(\Sigma_\alpha)+
{{m_{\pi}^2}} x^2 \Tr \left[Q_RQ_R^\dagger\sin^2\alpha +
P_A^2\cos^2\alpha + P_S^2\right] + \ldots.
&&\hfil(\beta=1)
\end{eqnarray}
From (\ref{LeffPQ}) we can now read off the curvatures for the 
different multiplets of the (pseudo-)Goldstone modes.
We see that there is a true flat direction, $Q_I$, which
describes true massless Goldstones of the
diquark condensation phase. These fields are the phases (a single $U(1)_B$
phase for $N_f=2$ and a set of $U(1)_B\times SU(N_f)_V/Sp(N_f)_V$ phases
for other $N_f$) of the diquark condensate. As we shall see in the
next section, the linear derivative terms in the effective
Lagrangian mix $Q_I$ and $Q_R$, thus the actual true Goldstone
excitations are certain linear combinations of $Q_I$ and $Q_R$.


The degeneracies of the multiplets follow from the definitions
(\ref{PQproj}) and are given by
\begin{eqnarray}\label{NPPQQ}
N_{P_S} &=& {N_f(N_f + 1)\over 2};
\qquad
N_{P_A} = {N_f(N_f - 1)\over 2} - 1;
\nonumber\\
N_{Q_R} &=& N_{Q_I} = {N_f(N_f - 1)\over 2}.
\hskip 15em(\beta=1)
\end{eqnarray}
They correspond to representations of the group $Sp(N_f)$, which is
the residual symmetry remaining
intact after spontaneous breaking. 
The number of the flat directions $Q_I$ exactly matches the number
of the broken generators in $SU(N_f)_V\times U(1)_B\to Sp(N_f)_V$.

\subsubsection{$\beta=4$}

The derivation of the curvatures of the effective potential
${\cal L}_{\rm st}$ in the $\beta=4$ case
follows  the same lines as in the $\beta=1$ case.
All the differences stem from the fact that $\Sigma$
is a symmetric unitary matrix in this case. Consequently, $\Sigma_c$ and
$\Sigma_d$ are given by (\ref{sigmacadj}) and (\ref{sigmadadj})
instead of (\ref{sigmac}) and (\ref{sigmad}).
In particular, the generator which rotates $\Sigma_c$ into $\Sigma_d$
is now given by
\begin{equation}\label{x2adj}
X_2 = -i\Sigma_d\Sigma_c^\dagger = \mat 0&1\\1&0 \emat,
\hfil \qquad (\beta=4)
\end{equation}
and the generator $X_1$, into which $B$ rotates and which is 
responsible for the splitting of both $P$ and $Q$ branches (in the
block decomposition (\ref{PiPQ})),
is given by
\begin{equation}\label{x1adj}
X_1 = iBX_2 = \mat 0&i\\-i&0 \emat.
\hfil \quad (\beta=4)
\end{equation}
It is now straightforward to rewrite the quadratic part of the
Lagrangian (\ref{L_exp})
in terms of the $PQ$-block decomposition. The mass matrix 
becomes diagonal in terms of the following
projections of $P$ and $Q$:
\begin{eqnarray}\label{PQprojadj}
P_S = \frac12(P + P^T) 
\quad\mbox{ and }\quad 
P_A = \frac12(P - P^T);\nonumber\\
Q_R = \frac12(Q + Q^\dagger )
\quad\mbox{ and }\quad 
Q_I = \frac12(Q - Q^\dagger).
&&\hfil (\beta=4)
\end{eqnarray}
The static part of the effective Lagrangian in terms of these
fields is given by
\begin{eqnarray}\label{LeffPQadj}
{\cal L}_{\rm st}(\Sigma)={\cal L}_{\rm st}(\Sigma_\alpha)+
{{m_{\pi}^2}} x^2 \Tr \left[Q_RQ_R^\dagger\sin^2\alpha +
P_S^2\cos^2\alpha + P_A^2\right] + \ldots,
&&\hfil (\beta=4)
\end{eqnarray}
which is similar to (\ref{LeffPQ}), but $S$ and $A$ labels 
exchange places. 


The main difference is in the dimension of the degenerate
multiplets which matches the representations of the
corresponding residual symmetry group. By inspection one finds
\begin{eqnarray}\label{NPPQQadj}
N_{P_A}& =& {N_f(N_f - 1)\over 2};
\qquad
N_{P_S} = {N_f(N_f + 1)\over 2} - 1;
\nonumber\\
N_{Q_R}& =& N_{Q_I} = {N_f(N_f + 1)\over 2}.
\hskip 15em(\beta=4)
\end{eqnarray}


\section{Spectrum}
\label{sec:spec}

In order to complete our program and determine 
the spectrum of low-lying excitations we must
take into account the derivative terms
in the Lagrangian (\ref{L}). They contribute in a non-trivial way to the
quadratic form in the 
Goldstone fields. Due to the non-Lorentz invariant
nature of the system we study at finite $\mu$,
the dispersion laws\footnote{In Euclidean field theory the dispersion
relation is given by the poles of the propagator in the 
$E \equiv ip_0$-plane.}
do not have a simple form, $E^2 =
\mbox{\boldmath $p$}^2 + m^2$.
The mass, as measured on the lattice and
given by the exponential fall-off of the propagator at
large Euclidean time, 
is the value of $E$, i.e. $ip_0$, 
at the pole of the propagator at {\boldmath $p$}$=0$. 
These pole masses, or rest energies, we shall now evaluate.

\subsection{Normal phase: $\mu<m_\pi/2$}

Expanding the derivative terms in the Lagrangian (\ref{L})
in the same way as we expanded the static part in the
previous section we obtain
\begin{equation}
{\cal L}(\Sigma) = {\cal L}(\Sigma_c)+
\frac12\Tr\left\{\left(
\d_\nu\Pi - \mu [B_\nu,\Pi]
\right)^2
+ m_\pi^2 \Pi^2
\right\}
+ \ldots,
\end{equation}
where $B_\nu = B\delta_{\nu0}$, as defined earlier in Section~\ref{sec:muloc}.
Using the Fourier decomposition of $\Pi$,
\begin{equation}
\Pi(x) = \sum_p \Pi_p \exp(-ipx) 
= \sum_p \Pi_p \exp(-Ex_0 - i\mbox{\boldmath$px$}),
\end{equation}
 we find that the dispersion law has the generic form
\begin{equation}
(E+b\mu)^2 = \mbox{\boldmath$p$}^2 + m_\pi^2,
\qquad\mbox{ or }\qquad
E = -b\mu + \sqrt{\mbox{\boldmath$p$}^2 + m_\pi^2},
\end{equation}
where $b$ is the baryon charge of the given excitation.
This is in agreement with the fact that the effect of $\mu$
on each state is simply an energy shift by $-b\mu$.
In particular, the dispersion law of the $P$-type Goldstone modes,
which carry no baryon charge,
are not affected by $\mu$ at all. The rest energies of the diquarks
$Q$ and antidiquarks $Q^\dagger$ are shifted according to their
charges. To summarize,
the dispersion laws in the normal phase 
are given by
\footnote{Note that though the mass of the diquarks $Q$
vanishes at the transition point, $\mu=m_\pi/2$, the dispersion
law is not linear, but quadratic, $E\sim \mbox{\boldmath$p$}^2$.
This is related to the well-known critical slowing down at a second order
phase transition.}
\begin{eqnarray}\label{dPQ}
P &:& E = \left(m_\pi^2 +\mbox{\boldmath$p$}^2\right)^{1/2} ;
\nonumber\\
Q^\dagger &:&  E = \left(m_\pi^2 +\mbox{\boldmath$p$}^2\right)^{1/2}
+ 2\mu;
\nonumber\\
Q &:&  E = \left(m_\pi^2 +\mbox{\boldmath$p$}^2\right)^{1/2}
- 2\mu.
\end{eqnarray}

The masses of these excitations as well as their degeneracies are given in
Tables \ref{tab:pq} and \ref{tab:pqadj}. A schematic picture of the mass
dependence is shown in Figures \ref{fig:spec} and \ref{fig:specadj}.

\subsection{Diquark condensation phase: $\mu>m_{\pi}/2$.}

In this phase the ground state changes and 
we need to expand around the rotated value of the
condensate $\Sigma=\Sigma_\alpha$.
We find%
\footnote{A useful observation which helps to write this in such a
compact form is that $\Tr[X_1,\Pi]\d_0\Pi=0$ for any $\Pi$.
This can be checked explicitly using the block decomposition of $\Pi$,
but it is easier to see that once you realize that the commutator
of two $X$-like generators is a $T$-like generator, and its
trace with another $X$-like generator, such as $\d_0\Pi$ is zero.}
\be
{\cal L}(\Sigma) = {\cal L}(\Sigma_\alpha)
&+& \frac12\Tr\left\{\left(
\d_\nu\Pi - \mu [V_\alpha B_\nu V_\alpha^\dagger,\Pi]
\right)\left(
\d_\nu\Pi - \mu [V_{\alpha}^\dagger B_\nu V_{\alpha},\Pi]
\right)
\right.\nonumber\\
&+& \left. m_\pi^2 \Pi^2\cos\alpha
\right\}
+ \ldots.
\ee
Expanding the product we find
\begin{equation}
{\cal L}(\Sigma) = 
\frac12\Tr\left\{\left(\d_\nu\Pi\right)^2
-2\mu\cos\alpha[B,\Pi]\d_0\Pi
\right\}
+{\cal L}_{\rm stat}(\Sigma)
+ \ldots
\end{equation}
with the static part given by (\ref{L_exp}). We observe, first of all,
that the linear derivative term contains only the charged fields, i.e. $Q$ and
$Q^\dagger$. The dispersion laws of the $P$ fields remain unaffected 
by the linear term. It remains Lorentz invariant, with  mass
given by the curvature of the static part of the Lagrangian.
In order to determine the
dispersion laws for the $Q$ and $Q^\dagger$ fields we need
to solve a secular equation obtained by substituting
Fourier decomposition of $Q$'s into
\be
{\cal L}(\Sigma)  &=& {\cal L}(\Sigma_\alpha)
+ \Tr\left\{\left(\d_\nu Q_R^\dagger\d_\nu Q_R
+\d_\nu Q_I^\dagger\d_\nu Q_I\right)
\right.\nonumber\\&&\left.
-4\mu\cos\alpha \left(Q_I^\dagger\d_0Q_R + Q_R^\dagger\d_0 Q_I\right)
+ 4\mu^2 Q_R^\dagger Q_R\sin^2\alpha
\right\} + \ldots,
\ee
where we only wrote the $Q$-dependent terms. This expression is
the same for either $\beta=1$ or $\beta=4$; only the definition of
$Q$'s is different: eqs. (\ref{PQproj}) or eqs. (\ref{PQprojadj}).
The secular equation has the form
\begin{equation}
\det\mat E^2-\mbox{\boldmath$p$}^2&-4\mu E\cos\alpha\\
-4\mu E\cos\alpha&E^2-\mbox{\boldmath$p$}^2 - 4\mu^2\sin^2\alpha\emat=0.
\end{equation}
We see that $Q_R$ and $Q_I$ are mixed by the linear derivative term.
The mixing is maximal when $\alpha=0$,  and vanishes when $\alpha=\pi/2$,
i.e. when $\mu\gg m_\pi$. For any $\alpha$ there is a solution
for which the rest energy $E(0)=0$ ---
 the Goldstone boson branch, which we denote $\tilde Q$.
For $\alpha=0$ it is given by $\tilde Q=Q=Q_R+Q_I$ (at rest,
$\mbox{\boldmath $p$} =0$). 
For $\alpha=\pi/2$ it is entirely $Q_I$. The other solution of the secular
equation, the linear combination of $Q_R$ and $Q_I$ orthogonal to $\tilde Q$,
is massive. We denote it by  $\tilde Q^\dagger$.

The dispersion laws for the meson multiplets $P_S$, $P_A$ and the 
mixed diquark-antidiquark multiplets $\tilde Q$ and $\tilde Q^\dagger$
are given by (in the $\beta=1$ case) the following relations:
\begin{eqnarray}\label{dPPQQ}
P_S &:& E^2 =  \mbox{\boldmath$p$}^2 + m_\pi^2x^2 ;
\nonumber\\
P_A &:&  E^2 =  \mbox{\boldmath$p$}^2 + m_\pi^2x^2\cos^2\alpha ;
\nonumber\\
\tilde Q^\dagger &:& E^2=\mbox{\boldmath$p$}^2 + 2 \mu^2 (1+3 \cos^2\alpha) +
2 \mu \sqrt{\mu^2 (1+3 \cos^2\alpha)^2+4  \mbox{\boldmath$p$}^2 \cos^2\alpha};
\nonumber\\
\tilde Q  &:& E^2 = \mbox{\boldmath$p$}^2 +2 \mu^2 (1+3 \cos^2\alpha) -
2 \mu \sqrt{\mu^2 (1+3 \cos^2\alpha)^2+4  \mbox{\boldmath$p$}^2
\cos^2\alpha}. \hskip 1em(\beta=1)  
\end{eqnarray}
The only difference in the dispersion laws for $\beta=4$ is the 
interchange of $P_S$ and $P_A$, 
and also that the degeneracy of the multiplets as given by
(\ref{NPPQQ}) and (\ref{NPPQQadj}) is different. Note that the dispersion
relation for the lowest lying 
mode, i.e. $\tilde Q$, is linear: $E \sim  p$. The slope is
a function of $\mu$: it vanishes at the transition point,
$x=1$, and approaches 1 for large $\mu$. The linear slope of low
energy excitations is characteristic of superfluidity.

The pole masses of the excitations, i.e. the position of the
pole at zero momentum,$E(0)$, 
are given in Table~\ref{tab:pq} for
$\beta=1$ and Table~\ref{tab:pqadj} for $\beta=4$.
The representations of the residual symmetry
groups are denoted by their Young diagrams. 
Because mesons are quark-quark or quark-antiquark pairs
they transform as rank two tensors (unless they are singlets). The
representations can be uniquely identified by their dimensions
found in (\ref{NPQ}), (\ref{NPPQQ}) and (\ref{NPQadj}), (\ref{NPPQQadj}). 
An explicit form of the representations is given in the Appendix. 
Figures~\ref{fig:spec} and ~\ref{fig:specadj} show schematic pictures
these results, together with the residual
symmetry groups in the different phases. In particular, one observes
that the spectrum is continuous at the transition point $x=1$.

\begin{figure}
\psfig{file=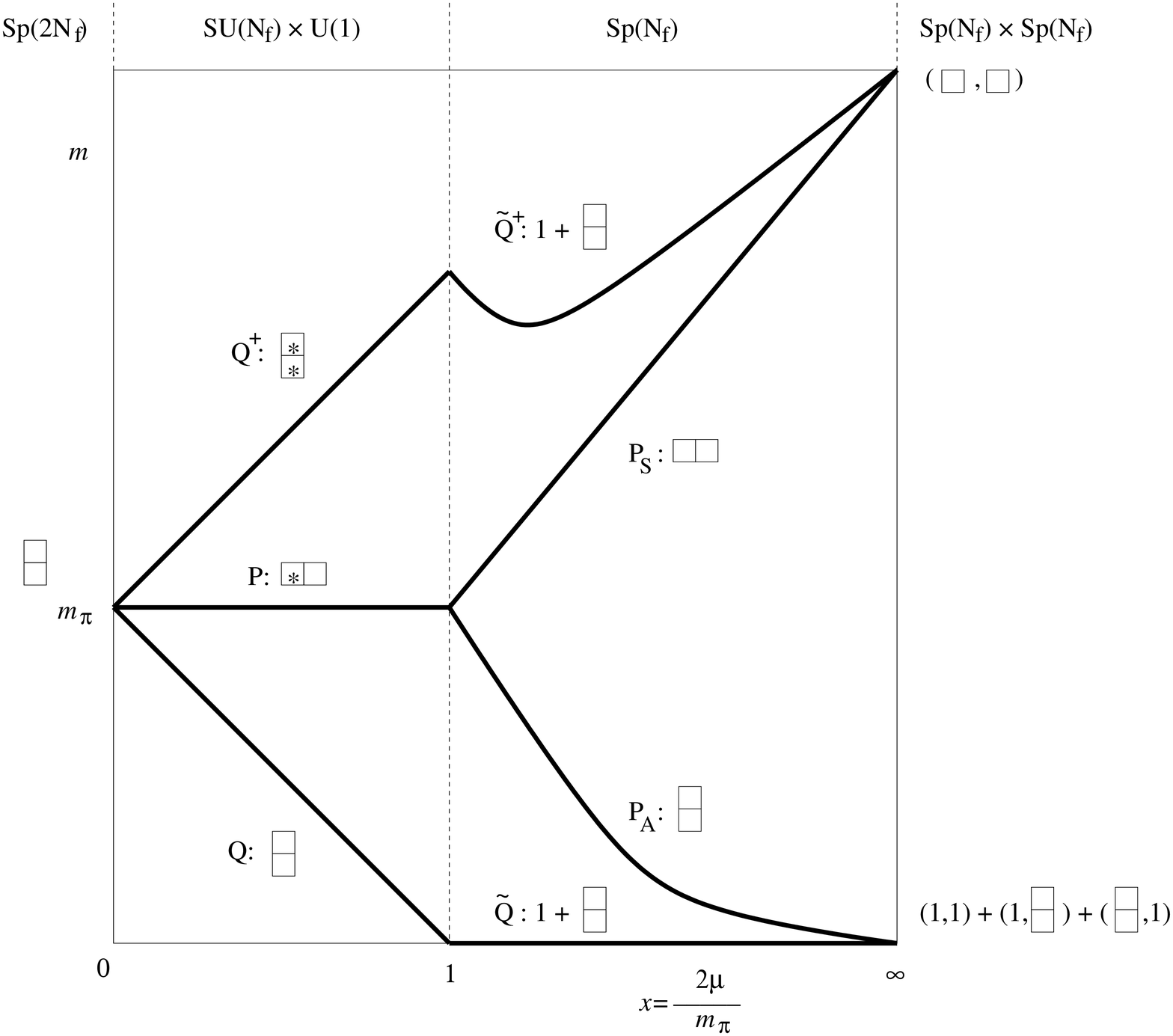,width=\textwidth}
\caption[]{Spectrum of two-color QCD ($\beta=1$) 
at finite $\mu$ and $m$ (schematic).
The branches are labeled according to (\ref{dPQ}), (\ref{PQproj}) and
\ref{PQprojadj}) 
and by Young's diagrams
showing the multiplet structure under the corresponding
residual symmetry. The star denotes the conjugate fundamental 
representation. The residual symmetry groups are 
marked above the plot. The degeneracies are given by (\ref{NPQ})
and (\ref{NPPQQ}).
For example, for the case of $N_f=2$ we find $N_P=3$, $N_Q=N_{Q^\dagger}=1$,
and $N_{P_S}=3$. The branch $P_A$ 
does not exist for $N_f=2$.
On the left/right of the plot the
residual symmetries and multiplet structure of the corresponding
limiting cases $\mu=0$ and $m=0$ are shown.
}
\label{fig:spec}
\end{figure}

\begin{table}
\caption[]{Spectrum (masses and degeneracies) of 
two-color QCD with fundamental quarks ($\beta=1$):
}
\vskip 1em

\renewcommand{\arraystretch}{2}
\centering

\begin{tabular}{|c||c|c|c||c|c|c|}
\hline
 & \multicolumn{3}{|c||}{$\mu <  m_\pi/2$: $SU(N_f) \times U(1)$} & 
\multicolumn{3}{|c|}{$\mu
  > m_\pi/2$: $Sp(N_f)$ } \\ \cline{2-7}   
& mass & degen. & rep. & mass & degen. &
rep.  \\
\hline
$P_S$ &  \raisebox{-.3cm}{{$m_\pi$}}  & 
\raisebox{-.3cm}{{$N_f^2-1$}} & $\hspace{.05cm}$
\raisebox{-.4cm}{* $\hspace{-.59cm}$
 \rule{.16mm}{.4cm} 
 $\hspace{-.28cm}$ \rule{.8cm}{.16mm} $\hspace{-1.085cm}$
 \rule[.4cm]{.8cm}{.16mm} $\hspace{-.68cm}$ \rule{.16mm}{.4cm} $\hspace{.11cm}$
 \rule{.16mm}{.4cm} } & {$2 \mu$}  &{ $\frac12N_f (N_f+1)$}   &
\rule{.16mm}{.4cm} $\hspace{-.28cm}$
 \rule{.8cm}{.16mm} 
 $\hspace{-1.085cm}$ 
 \rule[.4cm]{.8cm}{.16mm} $\hspace{-.68cm}$ \rule{.16mm}{.4cm} $\hspace{.11cm}$
 \rule{.16mm}{.4cm}  \\
\cline{1-1} \cline{5-7}
$P_A$ &  &  &  & {$m_\pi^2/2 \mu$} & {$\frac12N_f (N_f-1)-1$} &  
  \raisebox{-.22cm}{\rule{.16mm}{.8cm} 
 $\hspace{-.28cm}$ \rule{.4cm}{.16mm} $\hspace{-.68cm}$
 \rule[.4cm]{.4cm}{.16mm} $\hspace{-.68cm}$
 \rule[.8cm]{.4cm}{.16mm} $\hspace{-.28cm}$
 \rule{.16mm}{.8cm} } 
\\
\hline
$Q^\dagger$ &   $m_\pi+2 \mu$  & $\frac12 N_f (N_f-1)$ & $\hspace{.02cm}$
 \raisebox{-.22cm}{ 
 * $\hspace{-.48cm}$ \raisebox{.4cm}{*}
 $\hspace{-.45cm}$\rule{.16mm}{.8cm}   
 $\hspace{-.28cm}$ \rule{.4cm}{.16mm} $\hspace{-.68cm}$
 \rule[.4cm]{.4cm}{.16mm} $\hspace{-.68cm}$
 \rule[.8cm]{.4cm}{.16mm} $\hspace{-.28cm}$
 \rule{.16mm}{.8cm}  }   & {$2\mu\sqrt{1+3(m_\pi/2 \mu)^4}$} 
 & {$\frac12{N_f(N_f-1)}$} & 
 $1+$ \raisebox{-.22cm}{\rule{.16mm}{.8cm} 
 $\hspace{-.28cm}$ \rule{.4cm}{.16mm} $\hspace{-.68cm}$
 \rule[.4cm]{.4cm}{.16mm} $\hspace{-.68cm}$
 \rule[.8cm]{.4cm}{.16mm} $\hspace{-.28cm}$
 \rule{.16mm}{.8cm} }  \\
\cline{1-7}
$Q$ & $m_\pi-2\mu$ & $\frac12 N_f (N_f-1)$ & $\hspace{.04cm}$
 \raisebox{-.22cm}{\rule{.16mm}{.8cm}   
 $\hspace{-.28cm}$ \rule{.4cm}{.16mm} $\hspace{-.68cm}$
 \rule[.4cm]{.4cm}{.16mm} $\hspace{-.68cm}$
 \rule[.8cm]{.4cm}{.16mm} $\hspace{-.28cm}$
 \rule{.16mm}{.8cm} } &
{$0$}   & { $\frac12{N_f(N_f-1)}$}  & 
 $1+$ \raisebox{-.22cm}{\rule{.16mm}{.8cm} 
 $\hspace{-.28cm}$ \rule{.4cm}{.16mm} $\hspace{-.68cm}$
 \rule[.4cm]{.4cm}{.16mm} $\hspace{-.68cm}$
 \rule[.8cm]{.4cm}{.16mm} $\hspace{-.28cm}$
 \rule{.16mm}{.8cm} } \\
\hline 
\end{tabular}


\label{tab:pq}
\end{table}

\begin{figure}
\psfig{file=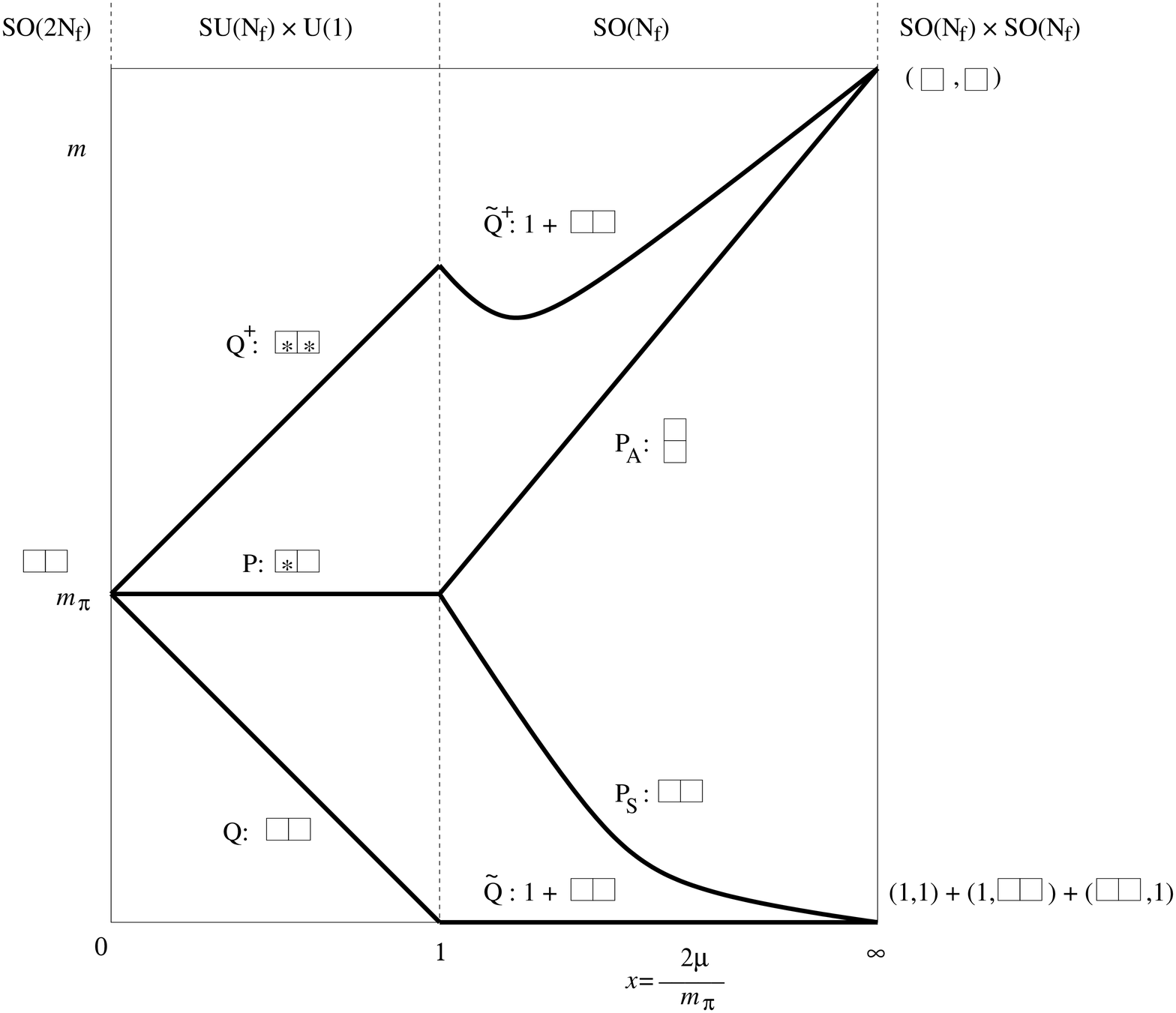,width=\textwidth}
\caption[]{Spectrum of {\em any}-color QCD with adjoint quarks
($\beta=4$) at finite $\mu$ and $m$ (schematic).
The branches are labeled according to (\ref{dPQ}), (\ref{PQproj}) and
(\ref{PQprojadj}) 
and by Young's diagrams
showing the multiplet structure under the corresponding
residual symmetry. The residual symmetry groups are 
marked above the plot. The degeneracies are given in Table \ref{tab:pqadj}.
For example, for the case of $N_f=2$: $N_P=3$, $N_Q=6$,
$N_{Q}=N_{Q^\dagger}=3$, $N_{P_S}=2$ and $N_{P_A}=1$.
On the left/right of the plot the
residual symmetries and multiplet structure of the corresponding
limiting cases $\mu=0$ and $m=0$ are shown.}
\label{fig:specadj}
\end{figure}

\begin{table}
\caption[]{Spectrum (masses and degeneracies) of 
any-color QCD with adjoint quarks ($\beta=4$).
}
\vskip 1em

\renewcommand{\arraystretch}{2}
\centering

\begin{tabular}{|c||c|c|c||c|c|c|}
\hline
 & \multicolumn{3}{|c||}{$\mu <  m_\pi/2$: $SU(N_f) \times U(1)$} & 
\multicolumn{3}{|c|}{$\mu
  > m_\pi/2$: $SO(N_f)$ } \\ \cline{2-7}   
& mass & degen. & rep. & mass & degen. &
rep.  \\
\hline
$P_A$ &  \raisebox{-.3cm}{{$m_\pi$}}  &
\raisebox{-.3cm}{{$N_f^2-1$}} & $\hspace{.08cm}$
\raisebox{-.4cm}{* $\hspace{-.59cm}$
 \rule{.16mm}{.4cm} 
 $\hspace{-.28cm}$ \rule{.8cm}{.16mm} $\hspace{-1.085cm}$
 \rule[.4cm]{.8cm}{.16mm} $\hspace{-.68cm}$ \rule{.16mm}{.4cm} $\hspace{.11cm}$
 \rule{.16mm}{.4cm} } & {$2 \mu$}  &{ $\frac12N_f (N_f-1)$}   &
  \raisebox{-.22cm}{\rule{.16mm}{.8cm} 
 $\hspace{-.28cm}$ \rule{.4cm}{.16mm} $\hspace{-.68cm}$
 \rule[.4cm]{.4cm}{.16mm} $\hspace{-.68cm}$
 \rule[.8cm]{.4cm}{.16mm} $\hspace{-.28cm}$
 \rule{.16mm}{.8cm} } 
\\
\cline{1-1} \cline{5-7}
$P_S$ &  &  &  & {$m_\pi^2/2 \mu$} & {$\frac12N_f (N_f+1)-1$} &  
\rule{.16mm}{.4cm} $\hspace{-.28cm}$
 \rule{.8cm}{.16mm} 
 $\hspace{-1.085cm}$ 
 \rule[.4cm]{.8cm}{.16mm} $\hspace{-.68cm}$ \rule{.16mm}{.4cm} $\hspace{.11cm}$
 \rule{.16mm}{.4cm}  \\
\hline
$Q^\dagger$ & $m_\pi+2 \mu$ & $\frac12 N_f (N_f+1)$ & $\hspace{.08cm}$
\raisebox{-.0cm}{* \hspace{-0.07cm} *
$\hspace{-0.99cm}$
 \rule{.16mm}{.4cm} 
 $\hspace{-.28cm}$ \rule{.8cm}{.16mm} $\hspace{-1.085cm}$
 \rule[.4cm]{.8cm}{.16mm} $\hspace{-.68cm}$ \rule{.16mm}{.4cm}
 $\hspace{.11cm}$ \rule{.16mm}{.4cm} }  & {$2\mu\sqrt{1+3(m_\pi/2 \mu)^4}$} 
 & {$\frac12{N_f(N_f+1)}$} & 
 $1+$ \raisebox{-.11cm}{\rule{.16mm}{.4cm} $\hspace{-.28cm}$
 \rule{.8cm}{.16mm} 
 $\hspace{-1.085cm}$ 
 \rule[.4cm]{.8cm}{.16mm} $\hspace{-.68cm}$ \rule{.16mm}{.4cm} $\hspace{.11cm}$
 \rule{.16mm}{.4cm}}  \\
\cline{1-7}
$Q$ & $m_\pi-2 \mu$ & $\frac12 N_f (N_f+1)$  & 
 \raisebox{-0cm}{ \rule{.16mm}{.4cm}  
 $\hspace{-.28cm}$ \rule{.8cm}{.16mm} $\hspace{-1.085cm}$
 \rule[.4cm]{.8cm}{.16mm} $\hspace{-.68cm}$ \rule{.16mm}{.4cm} $\hspace{.11cm}$
 \rule{.16mm}{.4cm} } &
{$0$}   & { $\frac12{N_f(N_f+1)}$}  & 
 $1+$ \raisebox{-.11cm}{\rule{.16mm}{.4cm} $\hspace{-.28cm}$
 \rule{.8cm}{.16mm} 
 $\hspace{-1.085cm}$ 
 \rule[.4cm]{.8cm}{.16mm} $\hspace{-.68cm}$ \rule{.16mm}{.4cm} $\hspace{.11cm}$
 \rule{.16mm}{.4cm}}  \\
\hline 
\end{tabular}


\label{tab:pqadj}
\end{table}

\section{Diquark source}
\label{sec:dqsource}

In three-color QCD the diquark condensate is not a color-singlet,
therefore, one cannot study the phenomenon of diquark condensation
by applying an external diquark source: such a source term
will not be gauge invariant. In the theories which we study in this
paper, e.g., in two-color QCD, the diquark condensate is
colorless. Therefore one can add a gauge-invariant source $j$
to the theory. This source plays a role similar to the role 
the quark mass $m$ plays with respect to the chiral condensate.
Such a non-zero source term $j$ is in fact used in lattice simulations
\cite{SU2lattice.new} in order to measure the value of the diquark
condensate in the limit $j\to 0$. In this section we show how
the results of the previous sections are modified in the presence of
a non-zero $j$.

\subsection{$\beta=1$}

Let us first rewrite the scalar diquark source term in the macroscopic theory
using our $SU(2N_f)$ spinor notations. In the notations of section 
\ref{sec:globsym} we find
\begin{eqnarray}\label{qqsource}
-i \frac j2 
\psi^TC\gamma_5\tau_2 I \psi + {\rm h.c.} 
=
- \frac j2 \Psi^T\sigma_2\tau_2 \mat iI&0\\0&iI\emat\Psi + {\rm h.c.}
\equiv - \frac 12 \Psi^T \sigma_2\tau_2  J \Psi + {\rm h.c.},
\nonumber
\end{eqnarray}
\begin{equation}
J = j\hat J\qquad{\rm and}\qquad 
\hat J = \mat iI&0\\0&iI\emat,
\hfil(\beta=1)
\end{equation}
the
antisymmetric matrix $I$ is defined in (\ref{I}), and the summation
 over the $N_f$ flavor indices has been suppressed. Comparing with the 
quark mass term (\ref{baremass}) we find that the two belong
to the same multiplet (i.e. they transform into one another) under the
$SU(2N_f)$ rotations.%
\footnote{There is a freedom in the choice of
the orientation of the diquark source $\hat J$. It is precisely
the same freedom as the one corresponding to the 
$SU(N_f)\times U(1)/Sp(N_f)$ degeneracy of the diquark condensation
vacuum $\Sigma_d$. An infinitesimal diquark source $j$ determines
the vacuum orientation of $\Sigma_d$. Our choices reflect this fact.
Indeed $\Sigma_d = \hat J^\dagger$.}
We can write the sum of the diquark source
and the bare mass term as
\begin{equation}
m\bar\psi\psi -i \frac j2 
(\psi^TC\gamma_5\tau_2 I \psi + {\rm h.c.}) 
= -\frac12 \Psi^T\sigma_2\tau_2 M_\phi\Psi, 
\hfil(\beta=1)
\end{equation}
where
\begin{equation}
M_\phi =  m \hat M + j\hat J = \sqrt{m^2+j^2} (\hat M\cos\phi +
 \hat J \sin\phi) = \sqrt{m^2+j^2}\hat M_\phi.
\end{equation}
We have used a notation similar to (\ref{osigma}) with
\begin{equation}\label{tanphi}
\tan\phi = \frac jm.
\end{equation}

It is easy to see how the diquark source term modifies the
effective Lagrangian (\ref{L}): we need to replace the mass matrix
$mG\hat M$ in (\ref{Lmu=0}) by
$\sqrt{m^2+j^2}G\hat M_\phi$.
The Gell-Mann$-$Oakes$-$Renner relation at $\mu=0$ becomes
\begin{equation}\label{mpij}
m_\pi^2 = {\sqrt{m^2+j^2}G\over F^2}.
\end{equation}
Using this fact we can write the effective Lagrangian for the
theory with both mass term and diquark source (compare to (\ref{L})) as
\begin{equation}\label{Leffj}
{\cal L}_{\rm eff}(\Sigma) = {F^2\over2} \left[
\Tr \nabla_\nu\Sigma \nabla_\nu\Sigma^\dagger 
- 2m_\pi^2\Re\Tr (\hat M_\phi\Sigma)
\right].
\end{equation}
Now we can repeat the steps we performed in section
\ref{sec:curv}, 
but for the
Lagrangian (\ref{Leffj}). 

Since the source term we introduced favors the direction of
$\Sigma_d$, we expect that
the minimum of the Lagrangian (\ref{Leffj}), $\overline\Sigma$, 
is again given by a linear combination
of $\Sigma_c$ and $\Sigma_d$ as in (\ref{osigma}),
\be\label{osigmaagain}
\overline{ \Sigma} = \Sigma_\alpha =\Sigma_c \cos\alpha
+\Sigma_d \sin\alpha = V_\alpha \Sigma_c V_\alpha^T, 
\quad {\rm where} \quad
V_\alpha = e^{i\alpha X_2},
\ee
and $X_2$ is the generator that rotates $\Sigma_c$ into $\Sigma_d$ as
before. The value of $\alpha$, as determined by the 
saddle-point equations, now depends on the value of the diquark source.
The proof 
of Section \ref{sec:globalminimum}
is not easily extendable to this case. We shall continue
on the assumption that the minimum {\em is} global.
We shall prove, however, that it is {\em a} minimum when we expand
to second order in fluctuations of $\Sigma$ and find
no linear terms and a positive quadratic form.
If we substitute this ansatz into the effective Lagrangian
(\ref{Leffj}) we find
\be
{\cal L}_{\rm st}(\Sigma_\alpha) = F^2m_\pi^2 N_f\left [ 
\frac{x^2}2 (\cos2\alpha-1) -2 \cos(\alpha -\phi)\right ].
\ee
The dependence of the angle $\alpha$ on $x$ follows from minimizing this 
Lagrangian and is given by
\begin{equation}\label{alphaphi}
x^2\cos\alpha\sin\alpha = \sin(\alpha-\phi),
\end{equation}
which is different from (\ref{alpha}), but coincides with it
in the limit $\phi\to0$, of course.
For finite $\phi$ the angle $\alpha$ is already non-zero
($\alpha=\phi$)
at $x=0$ as it should be, since a diquark source drives a non-zero
diquark condensate. At $x=\infty$, the value of $\alpha$
is $\pi/2$ independently of $\phi$. The major difference from
(\ref{alpha}) is that
the dependence of $\alpha$ on $x$ is analytic. There is no
phase transition as a function of $\mu$ when the diquark
source $j$ is nonzero. This is to be expected: the diquark
source plays the role an external magnetic field plays
in a ferromagnet. The diquark source {\em explicitly} breaks
$SU(N_f)_V\times U(1)_B$ symmetry down to $Sp(N_f)_V$
and there is only one phase (the one with residual $Sp(N_f)_V$ symmetry) 
for all values of $x$.

Our next step is to expand in powers of fluctuations
of $\Sigma$ around $\overline\Sigma$. This can be done 
following the algebra
in section \ref{sec:curv} with minimal modifications.
The Goldstone manifold is again parameterized by
\be
\Sigma= V_\alpha U \Sigma_c U^TV_\alpha^T,
\ee
but the second identity in (\ref{bsrotated}) is now given by
\be
V_{\alpha}^T \hat M_\phi V_{\alpha} 
= \left(V_{-\alpha} \hat M_\phi^\dagger V_{-\alpha}^T\right)^\dagger
= \Sigma_{\phi-\alpha}^\dagger
= \Sigma_c^\dagger
\cos(\alpha-\phi)
-\Sigma_d^\dagger \sin(\alpha-\phi),
\ee
whereas the first identity is the same as before. Using this algebra,
the effective Lagrangian to second order in the pion fields can be
expressed as
\be
\hskip -1em {\cal L}(\Sigma)&=&
{\cal L}_{\rm st}(\Sigma_\alpha)
+Fm_\pi^2 \left [ x^2 \cos\alpha \sin\alpha - \sin(\alpha -\phi)
\right ] {\rm Tr} (X_2 \Pi)
\nonumber\\
&+&
\frac12\Tr\left\{\left(\d_\nu\Pi\right)^2
-2\mu\cos\alpha[B,\Pi]\d_0\Pi
\right\} \nonumber \\
&+&
{{m_{\pi}^2}\over2} \left[
{x^2\over4}\left(\Tr[B,\Pi]^2\cos^2\alpha
-\Tr[X_1,\Pi]^2\sin^2\alpha \right)
+ \Tr \Pi^2\cos(\alpha-\phi)
\right] + \ldots
\label{lj}
\ee
Inspecting the linear terms we find that they vanish if $\alpha$
is chosen according to (\ref{alphaphi}). This means that
our ansatz (\ref{osigmaagain}) is indeed an extremum.

In order to find the mass spectrum we have to study the term of
second order in the $\Pi$-fields in the Lagrangian (\ref{lj}), and determine
the dispersion laws of the different (pseudo-)Goldstone fields.
Using block decomposition of the generators $\Pi$ (\ref{PiPQ}),
projections (\ref{PQproj}) of $P$ and $Q$ and the relationship
(\ref{alphaphi}) between $x$ and $\alpha$  we find
\begin{eqnarray} \label{Leffj1}
{\cal L}(\Sigma)&=&{\cal L}_{\rm st}(\Sigma_\alpha)
+ \Tr\left[\left(\d_\nu Q_R^\dagger\d_\nu Q_R
+\d_\nu Q_I^\dagger\d_\nu Q_I\right)
-4\mu\cos\alpha \left(Q_I^\dagger\d_0Q_R + Q_R^\dagger\d_0 Q_I\right)
\right]  \nonumber \\
&&+
{{m_{\pi}^2}} \Tr \left[
Q_IQ_I^\dagger{\sin\phi\over\sin\alpha}
+ 
Q_RQ_R^\dagger\left(x^2\sin^2\alpha + {\sin\phi\over\sin\alpha}\right)
\right] \nonumber\\ 
&&+
\Tr\left[\d_\nu P_A \d_\nu P_A+
P_A^2 m_\pi^2 \left(x^2\cos^2\alpha + {\sin\phi\over\sin\alpha}\right) \right]
\nonumber \\
&& + 
\Tr\left[\d_\nu P_S \d_\nu P_S+P_S^2 m_\pi^2 \left(x^2 +
    {\sin\phi\over\sin\alpha} \right) \right] + \ldots.  \hskip 2em (\beta=1) 
\end{eqnarray}

As in the case of a vanishing diquark source studied in the previous Section,
the linear derivative term 
contains only the charged fields $Q$ and $Q^\dagger$. Therefore the
dispersion relations for the neutral fields $P$ are not affected by this term:
the masses of the $P$ fields are given by the curvature of the static part of
the Lagrangian at the minimum.  
However the linear derivative term mixes the charged Goldstones. In order to
obtain the dispersion laws for the $Q$ fields, as in Section 10, we have to
solve a secular 
equation obtained by substituting the Fourier decomposition of these fields
into the effective Lagrangian (\ref{Leffj1}). The secular equation reads
\begin{equation} \label{seceqj}
\det\mat E^2-\mbox{\boldmath$p$}^2-m_\pi^2 \frac{\sin\phi}{\sin\alpha}&-4\mu
E\cos\alpha\\ 
-4\mu E\cos\alpha&E^2-\mbox{\boldmath$p$}^2 - 4\mu^2\sin^2\alpha -m_\pi^2
\frac{\sin\phi}{\sin\alpha}   \emat = 0. 
\end{equation}
 
The dispersion laws for the different Goldstone modes are therefore found to be
given by
\begin{eqnarray}\label{dPPQQj}
P_S &:& E^2 =  \mbox{\boldmath$p$}^2 + m_\pi^2 \left(
  x^2+\frac{\sin\phi}{\sin\alpha} \right)  ;
\nonumber\\
P_A &:&  E^2 =  \mbox{\boldmath$p$}^2 + m_\pi^2 \left(x^2\cos^2\alpha
  +\frac{\sin\phi}{\sin\alpha} \right) ;
\nonumber\\
\tilde Q^\dagger &:& E^2=\mbox{\boldmath$p$}^2 + m_\pi^2
  \frac{\sin\phi}{\sin\alpha} + 2 \mu^2 (1+3 \cos^2\alpha) \nonumber \\
 && \hskip 4em + 2 \mu \sqrt{\mu^2 (1+3 \cos^2\alpha)^2+4 \cos^2\alpha \; (
  \mbox{\boldmath$p$}^2+ m_\pi^2 \frac{\sin\phi}{\sin\alpha} )} ;
\nonumber\\
\tilde Q  &:& E^2=\mbox{\boldmath$p$}^2 + m_\pi^2
  \frac{\sin\phi}{\sin\alpha} + 2 \mu^2 (1+3 \cos^2\alpha) \nonumber \\
 && \hskip 4em - 2 \mu \sqrt{\mu^2 (1+3 \cos^2\alpha)^2+4 \cos^2\alpha \; (
  \mbox{\boldmath$p$}^2+ m_\pi^2 \frac{\sin\phi}{\sin\alpha} )}; \hskip
  1em(\beta=1)  ,
\end{eqnarray}
where $\alpha$ is an implicit function of $x$ and $\phi$ given by
(\ref{alphaphi}). Note that the $\tilde Q$ are no longer true 
Goldstone modes,
since the symmetry, which is broken spontaneously in the
diquark condensation phase $x>1$ at $j=0$, is now broken explicitly by
the diquark source term. The masses of the different multiplets are given by
the value of $E$ at  $\mbox{\boldmath$p$}=0$. The positivity of all masses
shows that the minimum given by the saddle-point equation (\ref{alphaphi}) is
at least a local minimum. This mass spectrum at a small non-zero diquark source
is depicted in Fig.~\ref{fig:spec/j}.

\subsection{$\beta=4$}

In the $\beta=4$ case the diquark source term has the form%
\footnote{As before, our previous choice of $\Sigma_d$
corresponds to our choice of $J$, i.e., $\Sigma_d=\hat J^\dagger$.}
\begin{eqnarray}\label{qqsource4}
i \frac j2 \psi^TC\gamma_5  \psi + {\rm h.c.}
=
\frac j2 \Psi^T\sigma_2 \mat i&0\\0&i\emat\Psi + {\rm h.c.}
\equiv - \frac 12 \Psi^T \sigma_2  J \Psi + {\rm  h.c.},
\nonumber
\end{eqnarray}
\begin{equation}
J = j\hat J\qquad{\rm and}\qquad 
\hat J = -\mat i&0\\0&i\emat
\hfil(\beta=4)
\end{equation}
The analysis follows the same lines as in the $\beta=1$ case.
We again introduce the combined mass matrix
\be
M_\phi = m \hat M + j \hat J = \sqrt{m^2+j^2} \hat M_\phi,
\ee
with $\hat M$ and $\hat J$ defined for $\beta =4$ in (\ref{mmm4}) and
(\ref{qqsource4}). 
In the expansion of the Lagrangian to second order in the $\Pi$-fields
for $\beta = 1$ we have only used the general commutation properties of
$\hat M_\phi$ which, with a proper redefinition of the generators
$X_1$ and $X_2$, are the same in the case $\beta =4$. We thus obtain
the same second order Lagrangian (\ref{lj}) as for $\beta=1$ resulting
in the same saddle point equation (\ref{alphaphi})
and the same mean field value of the
free energy. However, the mass spectrum is slightly different. We express the
quadratic part of the Lagrangian in terms of the block-decomposition
(\ref{PiPQ}) and the  
projections (\ref{PQprojadj}). The final result is
\begin{eqnarray} \label{Leffj4}
{\cal L}(\Sigma)&=&{\cal L}_{\rm st}(\Sigma_\alpha)
+ \Tr\left[\left(\d_\nu Q_R^\dagger\d_\nu Q_R
+\d_\nu Q_I^\dagger\d_\nu Q_I\right)
-4\mu\cos\alpha \left(Q_I^\dagger\d_0Q_R + Q_R^\dagger\d_0 Q_I\right)
\right]  \nonumber \\
&&+
{{m_{\pi}^2}} \Tr \left[
Q_IQ_I^\dagger{\sin\phi\over\sin\alpha}
+ 
Q_RQ_R^\dagger\left(x^2\sin^2\alpha + {\sin\phi\over\sin\alpha}\right)
\right] \nonumber\\ 
&&+
\Tr\left[\d_\nu P_S \d_\nu P_S+
P_S^2 m_\pi^2 \left(x^2\cos^2\alpha + {\sin\phi\over\sin\alpha}\right) \right]
\nonumber \\ 
&& + 
\Tr\left[\d_\nu P_A \d_\nu P_A+P_A^2 m_\pi^2 \left(x^2 +
    {\sin\phi\over\sin\alpha} \right) \right] + \ldots.  \hskip 2em (\beta=4) 
\end{eqnarray} 

\begin{figure}
\psfig{file=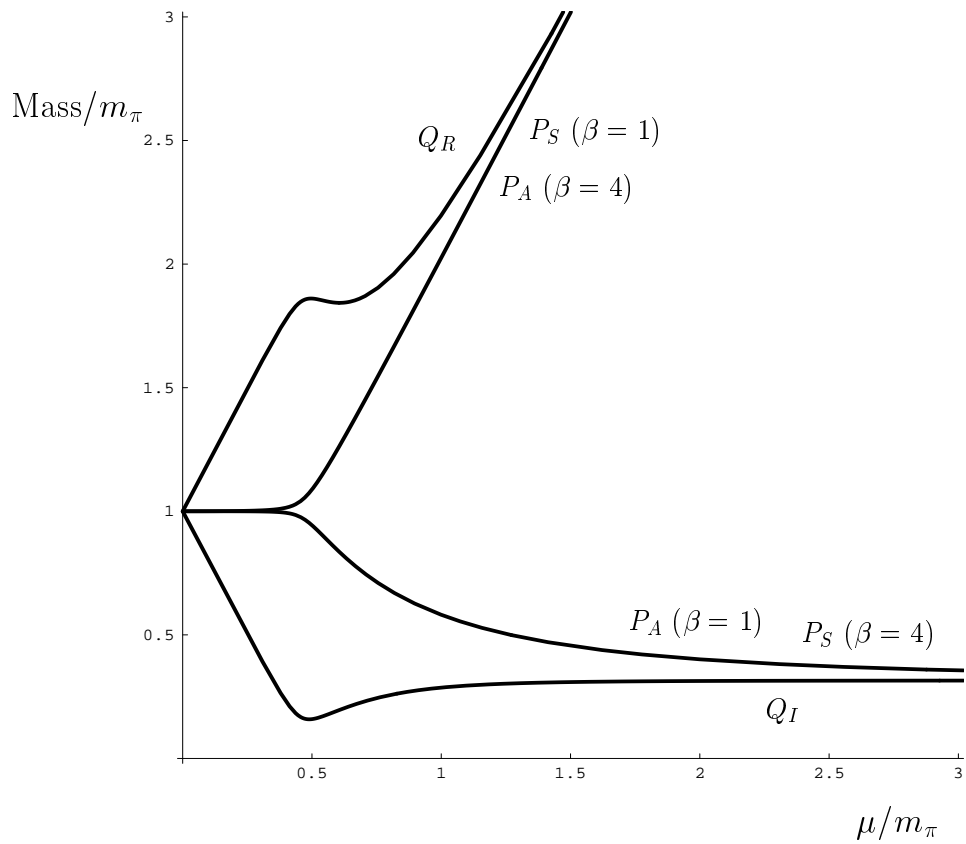,width=\textwidth}
\caption[]{Mass spectrum for a small non-vanishing source $j=0.1m$.}
\label{fig:spec/j}
\end{figure}

As in the $\beta=1$ case, the linear derivative term mixes the $Q$ and
$Q^\dagger$ fields. In order to get the dispersion relations of these fields,
the same secular equation as before (\ref{seceqj}) has to be solved. The
dispersion laws are the same as in (\ref{dPPQQj}) with the 
familiar exchange of labels $S\leftrightarrow A$, and the degeneracy of the
multiplets given by (\ref{NPPQQadj}) instead of (\ref{NPPQQ}).
The corresponding mass spectrum at a small non-zero $j$ is shown in
Fig.~\ref{fig:spec/j}.

\section{Condensate and baryon charge densities}
\label{sec:condensates}

In this section we derive the dependence
of the chiral condensate, the diquark condensate and
the baryon charge densities. These densities are obtained
by differentiating the vacuum energy density with respect
to $m$, $j$ and $\mu$,
\begin{equation}\label{derivatives}
\langle\bar\psi\psi\rangle = - {\d {\cal E}_{\rm
vac}\over\d m};
\qquad 
\langle\psi\psi\rangle = - {\d {\cal E}_{\rm
vac}\over\d j};
\qquad 
n_B =  -{\d {\cal E}_{\rm vac}\over\d \mu}.
\end{equation}
The shorthand symbol $\psi\psi$ denotes the l.h.s. of
the equation (\ref{qqsource}) in the $\beta=1$ case and of 
(\ref{qqsource4}) in the $\beta=4$ case without factor $j$.

In the effective theory the vacuum energy density is given by 
the value of the effective Lagrangian (\ref{L}) at the minimum,
\begin{eqnarray}
{\cal E}_{\rm vac} = {\cal L}_{\rm eff}(\Sigma_\alpha) 
&=& {F^2\over2}\left[ 
- 2\mu^2\Tr\left(\Sigma_\alpha B^T \Sigma_\alpha^\dagger B + BB\right)
- 
2m_\pi^2\Re\Tr \left(\Sigma_\phi^\dagger\Sigma_\alpha\right)
\right]
\nonumber\\&&
= -4N_fF^2\mu^2\sin^2\alpha - 2N_fG(m\cos\alpha +
j\sin\alpha),
\end{eqnarray}
where we used (\ref{tanphi}) and (\ref{mpij}).
Differentiating\footnote{The implicit dependence of $\alpha$ on $m$, $j$
and $\mu$ is not contributing because $\d{\cal E}_{\rm
vac}/\d\alpha=0$.}, we find
\begin{equation}\label{densities}
\langle\bar\psi\psi\rangle = 2N_fG \cos\alpha;
\qquad
\langle\psi\psi\rangle = 2N_fG \sin\alpha;
\qquad
n_B = 8N_fF^2\mu\sin^2\alpha,
\end{equation}
where the angle of the rotation of the condensate, $\alpha$, 
depends implicitly on $\mu$, $m$ and $j$ through the solution of
(\ref{alphaphi}). Note, in particular, that
the sum $\langle\bar\psi\psi\rangle^2+\langle\psi\psi\rangle^2$
does not depend on $\mu$, $m$ or $j$ (to the order in Chiral
Perturbation Theory we are working), which reflects 
the fact that the condensate rotates.

For $j=0$ the dependence of $\alpha$ on $\mu$ is simple:
$\alpha=0$, for $\mu<m_\pi/2$, and $\alpha=\arccos[m_\pi^2/(4\mu^2)]$
otherwise (\ref{alpha}). That means the densities (\ref{densities})
are constant for $\mu<m_\pi/2$, as they should
since the vacuum state does not change until $\mu$ reaches $m_\pi/2$.  The
results for the condensate and baryon charge densities are summarized in Table
3. 

\begin{figure}
\psfig{file=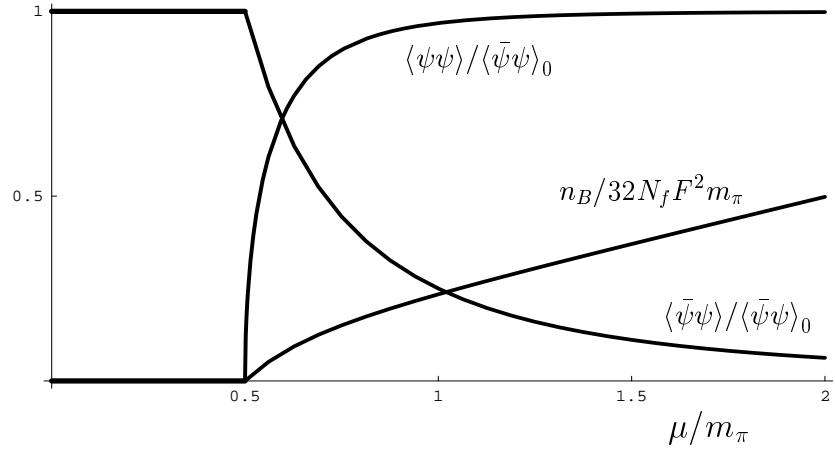,width=\textwidth}
\caption[]{The magnitudes of the chiral $\langle \bar{\psi}
\psi \rangle$ and the diquark $\langle
\psi \psi \rangle$ condensates in units of $\langle \bar{\psi}
\psi \rangle_0=2N_fG$ as a function of $\mu/m_\pi$ for zero diquark
source. Also the density of the baryon charge in units of $32N_fF^2m_\pi$
is shown.}
\label{fig:densities}
\end{figure}

\begin{figure}
\psfig{file=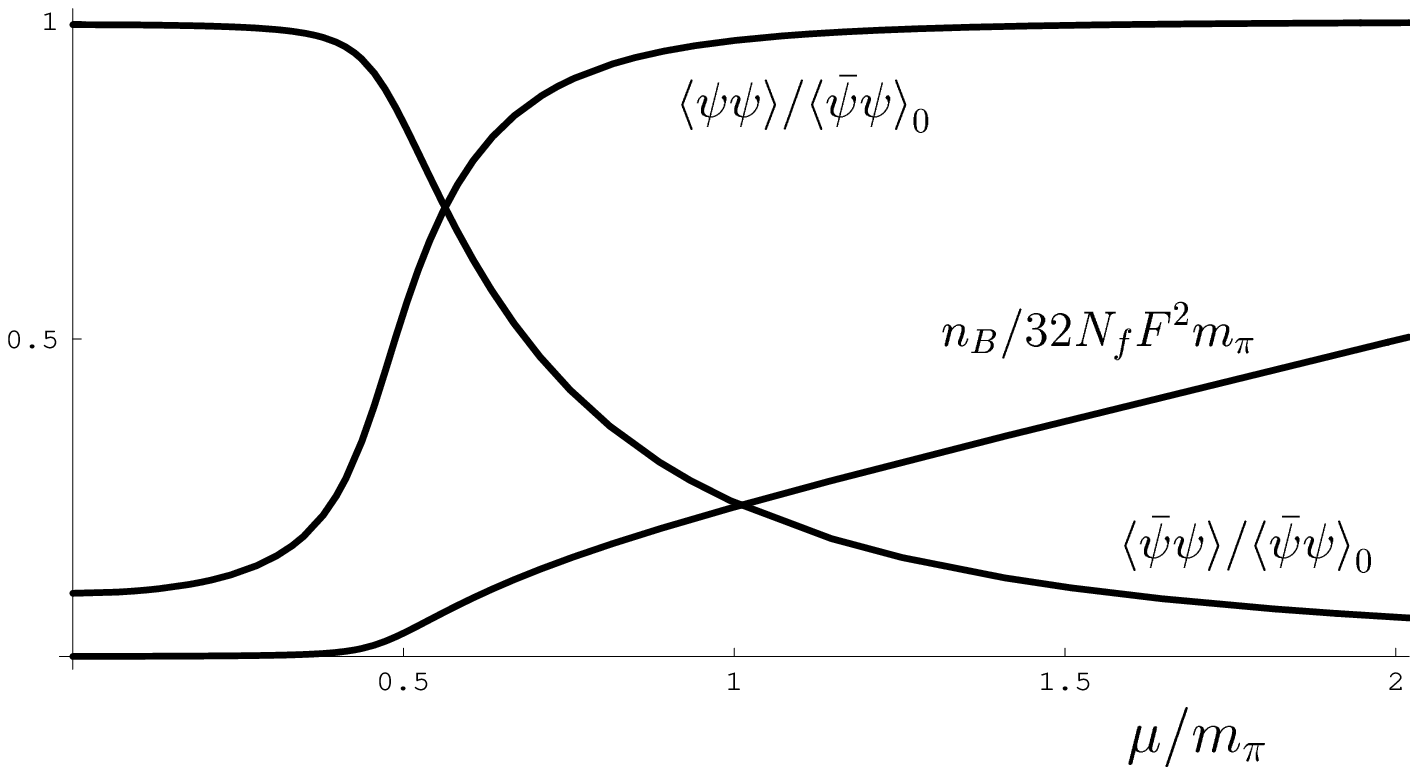,width=\textwidth}
\caption[]{The magnitudes of the chiral $\langle \bar{\psi}
\psi \rangle$ and the diquark $\langle
\psi \psi \rangle$ condensates in units of $2N_f\langle \bar{\psi}
\psi \rangle_0$ as a function of $\mu/m_\pi$ at small non-zero diquark
source $j=0.1m$.
Also the density of the baryon charge in units of $32N_fF^2m_\pi$
is shown.}
\label{fig:densitiesj}
\end{figure}

\begin{table}[h!]
\centering
\caption[]{The values of the chiral condensate, $\langle \bar \psi \psi
\rangle$, the diquark condensate, $\langle \psi \psi
\rangle$, and the baryon density $n_B$ in the two phases of the theory.}
\vskip 1em
\renewcommand{\arraystretch}{1.3}
\begin{tabular}{|c||c|c|c|}
\hline
phase  & $\langle \bar \psi \psi \rangle$ & $\langle  \psi \psi \rangle$& 
$n_B$\\
\hline
$\mu < m_\pi/2$ &  $\langle \bar \psi \psi \rangle_0$ & 0 & 0 \\
$\mu > m_\pi/2$ &  $\langle \bar \psi \psi \rangle_0
\left(\frac{m_\pi}{2\mu}\right)^2$ & 
$\langle \bar \psi \psi \rangle_0\sqrt{ 1 - \left(\frac{m_\pi}{2\mu}\right)^4}
  $ &  $ 8 \mu N_f F^2 \Big( 1-\left(\frac{m_\pi}{2\mu}\right)^4 \Big )$ \\
\hline
\end{tabular}
\renewcommand{\arraystretch}{1.0}
\end{table}
\vskip 1em
For $j \ne 0$ the angle $\alpha$ varies between  the values $\phi$ and
$\pi/2$ as a function of
$\mu$ and $m$, determined by the saddle point equation (\ref{alphaphi})
$x^2\sin\alpha \cos\alpha = \sin(\alpha -\phi)$. The saddle point is
a quartic equation which, in principle, can be solved analytically for
arbitrary $j$. 
The dependence of the densities (\ref{densities}) on $\mu$ is
shown in Fig. \ref{fig:densities} for $j =0$ and in Fig.
\ref{fig:densitiesj} for $j \ne 0$.

\section{Equation of state of the non-ideal diquark gas}
\label{sec:bose}

We now take a different look at the system
we study (quarks with two colors or any-color adjoint quarks) near the
threshold, $\mu\approx m_\pi/2$. First of all, we observe,
that for $\mu$ very close to $m_\pi/2$ the density of baryons is
very (arbitrarily) small. This means that we can describe the system
by an almost ideal gas. It should be a Bose gas, since the baryons,
i.e. diquarks, are bosons. The equation of state
of a dilute non-ideal 
Bose gas is a textbook problem \cite{Landau}. The interaction
between the bosons is crucial here (the dependence on its
strength is non-analytic, e.g. the attractive Bose gas is unstable).
The baryons in our system repel each other, as we shall see.
Moreover, we shall also see that the strength of this repulsion
is exactly the one that gives the correct equation of state
(\ref{densities}), (\ref{alpha}) at small densities:
\begin{equation}\label{n_Bsmall}
n_B = 32N_f F^2\left(\mu-{m_\pi\over2}\right) + \ldots.
\end{equation}
 This is a very non-trivial check of the consistency
of our effective Lagrangian!

Since we are working at very small densities, we can expand
the Lagrangian (\ref{Lmu=0}) up to next (quartic order) in the Goldstone
fields. The quartic terms will describe the interaction energy
of the Goldstones due to two-body scattering,
\begin{equation}\label{energy}
{\cal E}_{\rm int} =
- \frac1{24F^2}\Tr\left[\Pi,\d_\mu \Pi\right]\left[\Pi,\d^\mu \Pi\right]
- \frac{m_\pi^2}{24F^2}\Tr\Pi^4 + \ldots \ .
\end{equation}
We are now working in  Minkowski space and the extra minus sign
in front of the first term, compared to (\ref{Lmu=0}), is a
consequence of this.
The presence of the time derivatives should not confuse the reader.
Since we are dealing with a non-relativistic system, $\mu-m_\pi/2 \ll
m_\pi$, we only need the leading term in the time
dependence of the relativistic fields, i.e. $\d_0 \Pi\to \pm im_\pi
\Pi$. The spatial derivatives give only a subleading
contribution. The dominant effect is that of the $s$-wave amplitude.

We now consider the ground state of such a system with the
baryon density fixed (unlike before, when we fixed $\mu$).
Most of the particles will occupy the lowest energy, i.e., 
zero-momentum, level (the population
of the excited levels is the next order effect). These particles form a
condensate, whose amplitude we denote by $\Pi$. The energy of the
system is different from zero
only because of the interaction, and,  
as a function of $\Pi$, it is given to us by (\ref{energy}).
Looking at the block decomposition of $\Pi$ (\ref{PiPQ}) we
note that only $Q$ and $Q^\dagger$ (diquark and antidiquark)
fields carry non-zero baryon number. Therefore, we set $P=0$
in our case. Thus, from (\ref{energy}), we find for the dependence of the
energy on the magnitude of the $Q$ condensate
\begin{equation}\label{evacQ}
{\cal E}_{\rm int} = {m_\pi^2\over4F^2} \Tr(QQ^\dagger)^2 + \ldots,
\end{equation}
where we neglected subleading non-relativistic corrections.

Now we need to calculate $n_B$ as a function of $Q$.
Since $Q$ carries baryon charge 2, we can write%
\footnote{Another way to see (\ref{n_BQ})
is to use the fact that $n_B=-\d {\cal L}/\d\mu$. Thus,
from (\ref{L}) we find that 
$n_B=2F^2 \Tr B\Sigma^\dagger\partial_0\Sigma = 
2i\Tr B\Pi\partial_0\Pi$, which coincides with (\ref{n_BQ}).
}
\begin{equation}\label{n_BQ}
n_B = 2\,\Tr(Q^\dagger i\partial_0 Q - Qi\partial_0 Q^\dagger) = 
4 m_\pi \Tr QQ^\dagger + \ldots,
\end{equation}
where we neglected subleading corrections. Since all flavors
of diquarks are equivalent\footnote{There is only one flavor of diquarks for
  $\beta=1$, $N_f=2$. 
For larger $N_f$ the flavor $SU(N_f)$ symmetry ensures the
equivalence of all diquarks.}
the matrix $Q$ is proportional to $I$ for 
$\beta=1$, or to $1$ for $\beta=4$. Thus we can
relate traces in (\ref{evacQ}) and (\ref{n_BQ}) by $\Tr(QQ^\dagger)^2 =
(\Tr QQ^\dagger)^2/N_f$. Therefore, we find for the dependence  on
$n_B$ of the vacuum energy at fixed $n_B$,
\begin{equation}
{\cal E}_{\rm vac} = {n_B\over 2} m_\pi + {\cal E}_{\rm int} 
= {n_B\over 2} m_\pi + {n_B^2\over 64F^2N_f} + \ldots\ .
\end{equation}
The first term is the rest energy of the diquarks.
This determines the equation of state
\begin{equation}\label{n_Bsmall2}
\mu \equiv {\d {\cal E}_{\rm vac}\over \d n_B}
= {m_\pi\over2} + {n_B\over 32F^2N_f} + \ldots\ .
\end{equation}
The fact that the equations of state (\ref{n_Bsmall}) and
(\ref{n_Bsmall2}) are identical is a very nontrivial property of
the effective Lagrangian (\ref{L}). It is intimately related
to the symmetries (global and local) which determine the form of this
Lagrangian.

This calculation also shows that the diquarks Bose-condense \cite{EdCam}, 
and thus form a superfluid. Since the diquarks are charged we can call this
phase a superconducting phase.

\section{Conclusions and discussion}
\label{sec:conclusions}

In this paper we have derived the low-energy effective Lagrangian for
certain QCD-like theories at finite chemical potential $\mu$. It contains
the same number of phenomenological parameters as at $\mu=0$: the pion decay 
constant $F$ and the vacuum value
of the chiral condensate $\langle\bar\psi\psi\rangle_0$.
Using this effective Lagrangian we have completely determined its low-energy
properties. In particular, we have established the dependence of the ground
state, the condensates, and the masses of the
excitations on the chemical potential $\mu$, 
the bare quark mass $m$ and the diquark source $j$. 
The theories to which our analysis applies include: (i) two-color QCD with
quarks in fundamental color representation, and (ii) any-color QCD
with quarks in adjoint color representation. The unifying feature of these
theories is that the low-energy excitations, i.e. the Goldstone
bosons of spontaneously broken symmetries, include baryons in the
form of diquark states. Since the chiral Lagrangian
describes the dynamics of such low-energy excitations, we
can use the effective theory to describe phenomena associated with the
condensation of diquarks.

Since we are working only to the lowest order in Chiral Perturbation
Theory, we must remember that our approach has a limited
region of applicability. First of all, the momenta of the particles,
described by the Lagrangian (\ref{L}) must be  smaller
than 
the scale $M$ of the masses of the non-Goldstone
excitations, such as vector mesons, for example. In other words, the expansion
parameter of the chiral perturbation theory is $p/M$.
The bare quark mass is also included to 
leading order in $m$. It is obvious from (\ref{L}) that $m\sim
m_\pi^2\sim  p^2$ in the momentum power counting. The chemical
potential $\mu$ is an external parameter with the dimension of mass.
The interesting behavior of our theory is at values of 
$\mu \sim m_\pi \sim p$, and $ \mu$ is counted as order $p$ in the
momentum power counting.
Our effective theory thus applies as long as $m_\pi$ and $\mu$
are much smaller than $M$.

Our effective theory predicts a phase transition at a  value of
$\mu=m_\pi/2$. It is important to note that this transition occurs 
{\em within} the domain of applicability of the effective theory, as long as
$m_\pi$ is small, as discussed above, 
unlike, for example, the transition in QCD with three colors.
The transition at $\mu=m_\pi/2$ is a robust prediction of our effective theory.
The existence of such a transition  is confirmed by lattice
Monte Carlo simulations \cite{SU2lattice.old,SU2lattice.new}.

The fact that our effective Lagrangian (\ref{L}) predicts 
the transition at a value of $\mu$ equal to $m_\pi/2$ is a
direct consequence of the symmetries of the theory.
The coefficient of the $\mu$-dependent term
in the effective Lagrangian, responsible for the transition, 
is not an independent phenomenological parameter, but is fixed
rigidly by the {\em local} symmetry \cite{KST}. This symmetry
corresponds to the conservation of the baryon charge (among
other generators of $SU(2N_f)$). This ensures that the
charge of a composite state in the effective theory
is an additive sum of the charges of the constituents
(non-renormalization). Indeed, the factor two
in the relation $\mu_0=m_\pi/2$ is the baryon charge
of the diquark. This is in agreement with 
the value of $\mu$, below which no transition can occur
at zero temperature, given by the minimum value of
the mass per baryon number among all baryons in the theory
(see, e.g.,  \cite{HaJa98}). In three-color QCD this
would correspond to a value of $\mu_0$ approximately
equal to 1/3 of the nucleon mass (offset by the binding energy of
nuclear matter).%
\footnote{We remind the reader that we measure the baryon
charge in units of the $U(1)_B$ charge of a single quark.}

Another way of looking at the $\mu$-dependence in the theory
is the following. In the microscopic theory, given by the
Lagrangian (\ref{LmicroB}), $\mu$ enters as 
(a timelike component of) an Abelian gauge potential. 
As such, it can be completely removed from the Lagrangian
by the time-dependent gauge transformation
\begin{equation}\label{mutrans}
\psi\to e^{\mu\tau}\psi
\qquad\mbox{ and }\qquad
\psi^*\to e^{-\mu\tau}\psi^*,
\end{equation}
where $\tau$ is the Euclidean time\footnote{ Note that, in Euclidean
formulation, the $\psi$ and $\psi^*$ are independent variables and the
global flavor symmetry group is in fact $Gl(2N_f)$ \cite{TV}.}.
However, this does not mean that the partition function does not depend 
 on $\mu$. Let us consider first the case
of finite temperature. In this case, the boundary conditions
in the Euclidean time direction for the quarks change after
(\ref{mutrans}): they are
no longer antiperiodic. The dependence on $\mu$ in the partition function
comes entirely from the boundary conditions on the fermion fields
\begin{equation}\label{bc}
\psi|_{\tau={1/ T}} = -e^{\mu/ T}\psi|_{\tau=0}, 
\qquad\mbox{ and }\qquad
\psi^*|_{\tau={1/ T}} = -e^{-\mu/  T}\psi^*|_{\tau=0}.
\end{equation}
The $\mu$-independent
Lagrangian with such boundary conditions is completely equivalent
to the Lagrangian (\ref{LmicroB}) and usual antiperiodic boundary conditions.
This fact can be conveniently used in lattice simulations.
The corresponding effective theory can also be defined with
all $\mu$-dependence in the boundary conditions. The boundary
conditions for a given effective (composite) bosonic 
field $\phi$ should then read
$\phi(1/T)=\exp(b\mu/T)\phi(0)$, where $b$ is the baryon charge
of the field $\phi$ \cite{Evans}. In the case of the matrix valued field
$\Sigma$ we have 
\begin{equation}
\Sigma|_{\tau={1/T}} = 
\exp\left({\mu\over T}B\right)\Sigma|_{\tau=0}\exp\left({\mu\over
    T}B^T\right), 
\end{equation}
where $B$ is the baryon charge matrix (\ref{pb0p}).

In the limit of zero temperature, our intuition suggests that
the dependence of the partition function 
on boundary conditions should weaken and disappear. This is, however,
not completely true for the boundary conditions such as (\ref{bc})
due to their singular nature in the limit $1/T\to\infty$.
There is indeed an interval of $\mu$: $-\mu_0 < \mu < \mu_0$, where
the partition function does not depend on $\mu$. However,
outside this interval, the singular nature of the boundary conditions
starts playing a role: a phase transition occurs and a
$\mu$-dependence appears. 

Why is the approach of this paper not directly applicable to
real three-color QCD? The main reason is that at $\mu=0$ the effective
theory described by the chiral Lagrangian does not contain excitations
with non-zero baryon number (pions do not carry baryon charge).
It is easy to see that, as a consequence, 
applying either the method of local symmetry
of Section \ref{sec:muloc} or the gauge transformation described in this
section, one finds no dependence on $\mu$ in the effective chiral
Lagrangian of QCD. Related to that is the fact that the value
of $\mu_0\approx m_{\rm nucleon}/3$ is large, and the approach
of this paper, based on an expansion in small $\mu$, will
not reach it. 

As we emphasized, the theories we considered
are related by the fact that the fermion representations are
pseudo-real. This is a consequence of the antiunitary
symmetries of the Dirac operator (\ref{beta=1}), (\ref{beta=4}). 
We identified
two such symmetry classes, distinguished by the Dyson
index $\beta=1$ and $\beta=4$. The effective theories for these
two classes of theories are very similar and dual, or complementary,
to each other, 
from the point of view of the residual global symmetries.
In the diquark  (finite density) phase, for
the $\beta=1$ theories the residual flavor symmetry is given by
$Sp(N_f)$, while for the $\beta=4$ case it is $SO(N_f)$.
The excitations form multiplets which correspond to
symmetric or antisymmetric second rank tensor representations
of these groups (see Figs \ref{fig:spec},\ref{fig:specadj} 
and Tables \ref{tab:pq},\ref{tab:pqadj}).
One can see that the $\beta=1$ and $\beta=4$ cases ``mirror''
each other with respect to $Sp(N_f)\leftrightarrow SO(N_f)$
and symmetric $\leftrightarrow$ antisymmetric.
On the other hand, the Dirac operator for three-color QCD
with fundamental quarks does not have any antiunitary symmetries, 
and the fermion representations are complex. This case falls into
the third remaining Dyson class with the index $\beta=2$.

It would be interesting to apply the approach of this paper to {\em lattice}
theories with pseudoreal fermions, for example to two-color QCD with
fundamental quarks. The symmetries of such theories are different from
their continuum counterparts and were analyzed in
\cite{SU2lattice.new}. 
In particular, the transition to continuum limit may turn our to be
nontrivial. This can be related to the antiunitary symmetry
of the lattice Dirac operator for staggered fermions in fundamental
representation
\begin{equation}
{\cal D}_{xy} = \frac12\sum_{\mu} \eta_{x,\mu} (U_{x,\mu}\delta_{x+\hat\mu,y} -
U^\dagger_{x-\mu,\mu}\delta_{x-\hat\mu,y}),
\end{equation}
where $\eta_{x,\mu}=(-1)^{x_1+\ldots+x_{\mu-1}}$, and $U$ are
$SU(2)$ color matrices. The antiunitary symmetry of this lattice
Dirac operator is given by
$\tau_2 {\cal D} = {\cal D}^*\tau_2$, or
$[{\cal D},\tau_2K] = 0$.
Since $(\tau_2K)^2=-1$, we conclude that such Dirac operator
belongs to the class $\beta=4$. However, in the continuum limit,
the Dirac operator must be in the class $\beta=1$. Such an observation
was also made in \cite{KogutStone,Teper,V,SU2lattice.new} 
from the point of view of global
symmetries and their breaking, where it was pointed out that it is
not yet known how the apparent pattern of $SU(2N_f)\to O(2N_f)$
breaking becomes $SU(2N_f)\to Sp(2N_f)$ in the continuum.
Even though the symmetry may modify many of the details
of our analysis, when it is applied to lattice theories,
some features should be robust. Such would include the
phase transition at $\mu=m_\pi/2$, the relation similar to
$\langle\bar\psi\psi\rangle^2+\langle\psi\psi\rangle^2={\rm const}$
between the chiral and the diquark condensates, the linear dependence
of $n_B$ on $\mu$ near the transition, the existence
of several branches in the spectrum, similar to Figures
\ref{fig:spec},\ref{fig:specadj},\ref{fig:spec/j}, and many other
qualitative features.

\vskip 1.5cm
\noindent{\bf Acknowledgements}
\vskip 0.5cm
S. Hands, H. Leutwyler, M.-P. Lombardo,  S. Morrison, E. Shuryak and
D.K. Sinclair are acknowledged for useful discussions. J.B.K. is supported in
part by the National Science Foundation, NSF-PHY96-05199. 
D.T. and J.J.M.V. are partially supported by the US DOE grant
DE-FG-88ER40388. D.T. is supported in part by
``Holderbank''-Stiftung and by Janggen-P\"ohn-Stiftung. A.Z. is supported, in
part, by the National Science and Engineering Research Council of Canada
(NSERC).

\section*{Appendix}

In this appendix we analyze the representations of the remaining
symmetry groups 
shown in Table \ref{tab:pq} for $\beta =1$. 
They are all different subgroups of $Sp(2N_f)$. 

Under the symmetry group $Sp(2N_f)$ the Goldstone fields $\Sigma$ transform as
\be
\Sigma \rightarrow V \Sigma V^T.
\ee
With $\Sigma$ parameterized as $ U \Sigma_c U^T$ (with $U\in SU(2N_f)$)
and $V \Sigma_c V^T = \Sigma_c$ (the $2N_f \times 2N_f$ 
antisymmetric unit matrix is denoted by $\Sigma_c$)
it follows that the generators transform according to
\be
X\Sigma_c \rightarrow V X V^{-1}\Sigma_c=V X\Sigma_c V^T\qquad {\rm or}
\qquad (X\Sigma_c)_{ij}
\rightarrow V_{ik}V_{jl}
(X\Sigma_c)_{kl}.
\ee

From the transposition relation (\ref{xsigma}) it follows that 
$(X\Sigma_c)^T = -X\Sigma_c$.
If the symmetry group is $Sp(2N_f)$ the generators transform according to
an antisymmetric rank two representation of
$Sp(2N_f)$. The degeneracy is thus given by $2N_f^2 -N_f -1$.
If the symmetry group is $SU(N_f) \times U(1)$ the symmetry transformation
is given by
\be
V = \mat U_1 &0 \\ 0 & U_1^* \emat,
\label{Vsym}
\ee
the $P$-type generators transform according to 
$P^T\rightarrow U_1 P^T U_1^{-1}$
and the $Q$-type generators as $Q \rightarrow U_1 Q U_1^T$. Since the $Q$
are antisymmetric they transform according to an antisymmetric rank two
representation of $SU(N_f)$. The fields $Q^\dagger$ transform according
to the conjugate representation. 
The dimension of both representations is equal to
$N_f(N_f-1)/2$. Since the $P$-type generators are traceless, the
degeneracy is given by $N_f^2 -1$.

In the diquark condensation phase the symmetry group $Sp(2N_f)$ is with
respect to the rotated antisymmetric unit matrix, $\Sigma_\alpha$, with
symplectic transformations defined by
\be
V \Sigma_\alpha V^T = \Sigma_\alpha.
\ee 
According to the argument at the beginning of this appendix, the
transformation properties of the generators are given by
\be
(X_\alpha\Sigma_\alpha)_{ij}
\rightarrow V_{ik}V_{jl}
(X_\alpha\Sigma_\alpha)_{kl},
\ee
where the $X_\alpha$ are the rotated generators defined by 
\be
X_\alpha = V_\alpha X V_\alpha^{-1},
\label{xalpha}
\ee
and $V_\alpha=\exp i\alpha X_2/2$ defined in (\ref{vsv}).  

With $Sp(N_f)$ as symmetry group  the symmetry transformations are given
by (\ref{Vsym}) with $U_1^* = -I U_1 I $ (the $N_f\times N_f$ antisymmetric
unit matrix is denoted by $I$). Since in this case $\{V, X_2 \} =0$
and using that $\Sigma = V_\alpha U \Sigma_c U^T V_\alpha^T$ (see
eq. (\ref{vusuv})) we find that the representations can be discussed
in terms of the fields $U\Sigma_c U^T$ with the familiar $PQ$-block
structure of the generators.  
The $P$-type generators
thus transform as $P^TI \rightarrow U_1 P^T I U_1^T $. The symmetric and
and antisymmetric components of $PI$ corresponding to $P_S I$ and $P_A I$,
respectively, transform independently. The dimensions of the representations
are given by $N_f(N_f+1)/2$ and $N_f(N_f-1) -1$, respectively.
The generators $Q$ and $IQ^\dagger  I$ transform in the same way, 
and thus the linear combinations $\tilde Q$ and $\tilde Q^\dagger$ transform
in the same way. Since $Q$ is antisymmetric they transform according to
an irreducible rank two representation with dimension equal to
$N_f(N_f-1)/2$. 

Finally, for symmetry group $Sp(N_f) \times Sp(N_f)$ the symmetry 
transformation
is given by
\be
V = \mat U_1 &0 \\ 0 & U_2 \emat,
\label{VsymSS}
\ee
with both $U_1$ and $U_2$ symplectic transformations. In this case it
is imperative to consider the rotated generators. However, we only
need the rotated generators for $\mu \rightarrow \infty$.
In terms of the 
block structure of the unrotated generators, we obtain after a
straightforward calculation for $\alpha = \pi/2$,
\be
X_\alpha \Sigma_\alpha =
\mat -IP_AI + iQ_I I & iIP_S +Q_R \\
     iP_SI +IQ_RI & P_A -iIQ_I \emat.
\label{rotdecom}
\ee
Therefore the combination $iIP_S + Q_R$ transforms as
\be
iIP_S + Q_R \rightarrow U_1 (iIP_S + Q_R) U_2^T.
\ee
Indeed, this transformation corresponds to the Young tableaux given
in  upper right corner of Fig. 1. Notice that for $\mu \rightarrow
\infty$ we have that $\tilde Q^\dagger \rightarrow Q_R$. The 11- and
22-blocks of the matrix (\ref{rotdecom}) correspond to two different
linear combinations of $P_A$ and $ Q_I$. One combination transforms
as a rank two tensor with respect to $U_1$, and the other combination
as a rank two tensor with respect to $U_2$. Since both combinations
are antisymmetric with respect to transposition, they transform
according to the Young tableaux given in the lower right corner of 
Fig. 1. The additional singlet terms arise because the irreducible
representations are traceless.

In case of $Sp(N_f)$ symmetry, $U_2 = U_1^* =-IU_1 I$, the
transformation
properties of $P_S$, $P_A$, $Q_R$ and $Q_I$ can be obtained from
(\ref{rotdecom}) by combining the transformation properties of
the diagonal blocks and of the off-diagonal blocks.  The results are 
in agreement with the discussion in the paragraph following 
eq. (\ref{xalpha}).

The same analysis can be performed for $\beta =4$. In fact, because
the matrix $ I$ is absent, this case is somewhat simpler, and we
leave it as an exercise to the reader.


\begin{thebibliography}{9}
\bibitem{Bielefeld}  See e.g.  Proceedings of {\it QCD at Finite Baryon 
        Density},
        Bielefeld, April 1998, F. Karsch and M.-P Lombardo, eds,
        Nucl. Phys. {\bf A 642}, 1998.
\bibitem{lattice.mu}
        J.B. Kogut, M.P. Lombardo and D.K. Sinclair,
                Phys. Rev. D 51 (1995) 1282;
                Nucl. Phys. B, Proc. Suppl. 42 (1995) 514;
        I.M. Barbour, S.E. Morrison, E.G. Klepfish,
                J.B. Kogut, M.P. Lombardo,
                Nucl. Phys. Proc. Suppl. {\bf 60A} (1998) 220.
\bibitem{SC1}
        D. Bailin and A. Love, Phys. Rept. {\bf 107} (1984) 325;
        M. Alford, K. Rajagopal and F. Wilczek,
                Phys. Lett. {\bf B422} (1998) 247,
                Nucl. Phys. {\bf B537} (1999) 443;
        R. Rapp, T. Sch\"afer, E.V. Shuryak and M. Velkovsky,
                Phys. Rev. Lett. {\bf 81} 1998 53; 
                Ann. Phys. 280 (2000) 35.
\bibitem{SU2lattice.old}
        E. Dagotto, F. Karsch, and A. Moreo,
        Phys. Lett. B {\bf 169} (1986) 421;
        E. Dagotto, A. Moreo, and U. Wolff, Phys. Rev. Lett. {\bf 57}
        (1986) 1292; Phys. Lett. B {\bf 186} (1987) 395.
\bibitem{SU2lattice.new}
        S. Hands, J.B. Kogut, M.-P. Lombardo,
        S.E. Morrison, Nucl. Phys. B 558 (1999) 327;
        S. Hands and S.E. Morrison, hep-lat/9902012, hep-lat/9905021.
\bibitem{Thomasmu}T. Sch\"afer, Phys. Rev. {\bf D57} (1998) 3950.
\bibitem{St96}
        M.A. Stephanov, Phys. Rev. Lett. {\bf 76} (1996) 4472;
        Nucl. Phys. Proc. Suppl. {\bf 53} (1997) 469.
\bibitem{KST}
        J. Kogut, M.A. Stephanov and D. Toublan, Phys. Lett. {\bf B464} (1999)
        183.
\bibitem{mulargeNc}
        R.F. Alvarez-Estrada and  A. Gomez Nicola,  Phys.Lett. {\bf
        B355} (1995) 288; errat. Phys.Lett. {\bf B380} (1996) 491.
\bibitem{DeTar}
        C.~DeTar, {\it Quark-gluon plasma in numerical simulations of
        QCD}, in {\it Quark gluon plasma 2}, R. Hwa ed., World
        Scientific 1995.
\bibitem{Smilref}
        A.V. Smilga, Phys. Rep. 291, (1997) 1.
\bibitem{pauli-gursey}
        W. Pauli, Nuovo Cimento {\bf 6} (1957) 205 ;
        F. G\"ursey, Nuovo Cimento {\bf 7} (1958) 411.
\bibitem{DuKo}
        S. Duane and J.B. Kogut, Nucl. Phys. {\bf B275} (1986) 398.
\bibitem{DKPR}
        S. Duane, A.D. Kennedy, B.J. Pendleton and D. Roweth,
        Phys. Lett. {\bf B195} (1987) 216.
\bibitem{SUSY} V.A. Novikov, M.A. Shifman, A.I. Vainshtein and V.I. Zakharov,
        Nucl. Phys. {\bf B260} (1985) 157.
\bibitem{VafaW}
        C. Vafa and E. Witten, Nucl. Phys. {\bf B234} (1984) 173.
\bibitem{Shifman-three}
        S. Dimopoulos, Nucl. Phys. {\bf B168} (1980) 69;
        M. Vysotskii, Y. Kogan and M. Shifman,
                Sov. J. Nucl. Phys. {\bf 42} (1985) 318;
        D.I. Diakonov and V.Yu. Petrov, Lecture notes in physics, {\bf
                417}, Springer 1993.
\bibitem{Peskin}
        M.E. Peskin, Nucl. Phys. {\bf B175} (1980) 197.
\bibitem{SmilV}
        A. Smilga and J.J.M. Verbaarschot, Phys. Rev. D51 (1995) 829.
\bibitem{Dyson} F.J. Dyson, J. Math. Phys. {\bf 3} (1962c) 1199.
\bibitem{SV}E.V. Shuryak and J.J.M. Verbaarschot,
        Nucl. Phys. {\bf A560} (1993) 306.
\bibitem{V}
        J.J.M. Verbaarschot, Phys. Rev. Lett. {\bf 72} (1994) 2531;
        Phys. Lett. {\bf B329} (1994) 351.
\bibitem{Zirnall}M. Zirnbauer, J. Math. Phys. {\bf 37} (1996) 4986.
\bibitem{Kyoto} J.J.M. Verbaarschot, {\it Lectures given at APCTP -
        RCNP Joint International School on Physics of Hadrons and QCD}, Osaka,
        Japan, 1998 and the {\it 1998 YITP Workshop on QCD
        and Hadron Physics}, Kyoto, Japan, 1998, hep-ph/9902394.
\bibitem{HOV}
        M.A. Halasz, J.C. Osborn and  J.J.M. Verbaarschot,
        Phys. Rev. {\bf D56} (1997) 7059.
\bibitem{Weinb} S. Weinberg, Physica {\bf A96} (1979) 327.
\bibitem{GaL}
        J. Gasser and H. Leutwyler, Ann. Phys. {\bf 158}, 142
        (1984); J. Gasser and H. Leutwyler, Nucl. Phys. B{\bf 250},
        465 (1985); H. Leutwyler, Ann. Phys. {\bf235} (1994) 165.
\bibitem{TV}
        D. Toublan and J.J.M. Verbaarschot, Nucl. Phys. {\bf B560} (1999) 259.
\bibitem{Fyodorov}  Y.V. Fyodorov, B.A. Khoruzenko and H.-J.
         Sommers, Phys. Lett. {\bf A226} (1997) 46; Y.V. Fyodorov, M. Titov
         and H.-J. Sommers, Phys. Rev. {\bf E58} (1998) 1195.
\bibitem{Efetov} K.B. Efetov, Phys. Rev. Lett. {\bf 79} (1997)
         491; Phys. Rev. {\bf B56} (1996) 9630; A.V. Kolesnikov and
         K.B. Efetov, Waves in Random Media {\bf 9} (1999) 71.
\bibitem{Landau}
        L.D. Landau, E.M. Lifshitz and L.P. Pitaevskii, {\it Statistical
        Physics, Part 2, v. 9}, Ch.25;  Oxford ; New York : Pergamon
        Press, 1980.

\bibitem{HaJa98}M.A. Halasz, A.D. Jackson, R.E. Shrock,
                M.A. Stephanov and J.J.M. Verbaarschot,
                Phys. Rev. {\bf D58} (1998) 096007.
\bibitem{EdCam}E. Shuryak, Lectures given at NATO Advanced Study 
Institute, in "Confinment, Duality and Nonperturbative Aspects of QCD",
Cambridge, 1997, 307.
\bibitem{Evans} T.S. Evans, hep-ph/9510298.
\bibitem{KogutStone}J.B. Kogut, H. Matsuoka, M. Stone, H.W. Wyld, S. Shenker,
  J. Shigemitsu and D.K. Sinclair, Nucl. Phys. {\bf B225} (1983) 93.  
\bibitem{Teper}S. Hands and M. Teper, Nucl. Phys. {\bf B347} (1990)
         819.  
\end{thebibliography}
\end{document}